\documentclass[useAMS,usenatbib]{mn2e}

\usepackage{graphicx}
\usepackage{color}

\newcommand{\ud}{\,\mathrm{d}}

\title[Galaxy group masses via caustic analysis]
	{Galaxy And Mass Assembly: Estimating galaxy group masses via caustic analysis}

\author[M. Alpaslan et al.]
	     {Mehmet Alpaslan$^{1,2}$, Aaron S.G.~Robotham$^{1,2}$, Simon Driver$^{1,2}$, Peder Norberg$^3$,\newauthor
	     John A.~Peacock$^4$, Ivan Baldry$^5$, Joss Bland-Hawthorn$^6$, Sarah Brough$^7$,\newauthor
	     Andrew M.~Hopkins$^8$, Lee S.~Kelvin$^{1,2}$, Jochen Liske$^7$, Jon Loveday$^9$,\newauthor
	     Alexander Merson$^3$, Robert C. Nichol$^{10}$, and Kevin Pimbblet$^{11}$\\
$^1$SUPA, School of Physics and Astronomy, University of St Andrews, North Haugh, St Andrews, Fife, KY16 9SS, UK\\
$^2$International Centre for Radio Astronomy Research, 7 Fairway, The University of Western Australia, Crawley, Perth, \\Western Australia 6009, Australia\\
$^3$Institute for Computational Cosmology, Department of Physics, Durham University, South Road, Durham, DH1 3LE, UK\\
$^4$Institute for Astronomy, University of Edinburgh, Royal Observatory, Edinburgh EH9 3HJ, UK\\
$^5$Astrophysics Research Institute, Liverpool John Moores University, Twelve Quays House, Egerton Wharf, Birkenhead, CH41 1LD, \\UK\\
$^6$Sydney Institute for Astronomy, University of Sydney, School of Physics A28, NSW 2006, Australia\\
$^7$European Southern Observatory, Karl-Schwarzschild-Str. 2, 85748 Garching, Germany\\
$^8$Australian Astronomical Observatory, PO Box 296, Epping, NSW 1710, Australia\\
$^9$Astronomy Centre, University of Sussex, Falmer, Brighton, BN1 9QH, UK\\
$^{10}$Institute of Cosmology and Gravitation, University of Portsmouth, Dennis Sciama Building, Burnaby Road Portsmouth, PO1 3FX, \\UK\\
$^{11}$School of Physics, Monash University, Clayton, Melbourne, Victoria 3800, Australia}

\date{Submitted 2012 March 28}

\pagerange{\pageref{firstpage}--\pageref{lastpage}}\pubyear{2011}

\begin{document}

\label{firstpage}

\maketitle

\begin{abstract}
We have generated complementary halo mass estimates for all groups in the Galaxy And Mass Assembly Galaxy Group Catalogue (GAMA G$^3$Cv1) using a modified caustic mass estimation algorithm, originally developed by \citet{diaferio_infall_1997}. We calibrate the algorithm by applying it on a series of 9 GAMA mock galaxy light cones and investigate the effects of using different definitions for group centre and size. We select the set of parameters that provide median-unbiased mass estimates when tested on mocks, and generate mass estimates for the real group catalogue. We find that on average, the caustic mass estimates agree with dynamical mass estimates within a factor of 2  in $90.8 \pm 6.1$ \%  groups and compares equally well to velocity dispersion based mass estimates for both high and low multiplicity groups over the full range of masses probed by the G$^3$Cv1.
\end{abstract}

\section{Introduction}

A quantitative understanding of the largest structures in the Universe provides a rigorous test of cosmology and dark matter simulations \citep{eke_cluster_1996}. One of the principal assumptions of the $\Lambda$CDM paradigm is that galaxy groups serve as tracers of the underlying dark matter haloes. It is thought that prior to the decoupling of matter and radiation,  dark matter particles, with their smaller interaction cross-section, formed overdense regions under the influence of gravity. Following  decoupling, baryonic matter is left to free fall into these overdense regions and form the building blocks of the large scale structure of the universe \citep{springel_simulations_2005}. Empirical measurements of the distribution and masses of galaxy groups therefore provide a powerful constraint for different dark matter models.  

Traditionally, mass estimates of groups are calculated virially, via measurements of the velocity dispersions of their members, e.g., \citet{hughes_mass_1989, carlberg_galaxy_1996,girardi_optical_1998, tucker_loose_2000}. By assuming that the group is in virial equilibrium, the dynamical mass of the group follows the relation $M \propto \sigma^2 R$. The obvious limitation of this method is that the relation will hold only out to the virial radius of the galaxy group, so mass estimates made using galaxies beyond this radius become less reliable. More accurate mass estimates at large radii are possible using weak lensing \citep{kaiser_method_1995}, however the obvious drawback of this approach is the observational challenge involved in measuring the lensing signal for a large galaxy group sample.

A different approach to group mass estimation is to look at the distribution of galaxies within a group in redshift space  (the projected distance $r$ from the group centre and the line-of-sight velocity $v$ with respect to the median group redshift for every member of the group) and estimate the group escape velocity by interpreting the distinct shape of this distribution. This method of analysing galaxy group members in redshift space was first introduced as a mass estimator by \citet{diaferio_infall_1997} and \citet{diaferio_mass_1999} (hereafter DG97 and D99) and is known as the caustic mass estimation technique. An early version of the method was first used in an attempt to constrain cosmological parameters, specifically $\Omega_M$ \citep{regos_infall_1989} with little success. In identifying that velocity measurements of galaxy groups were heavily affected by random  motions, as well as by comparing observed caustics to those predicted by cosmological models, DG97 determined that this method is unreliable for determining $\Omega_M$. \citet{navarro_universal_1997} went on to demonstrate that the density profiles of dark matter haloes do not vary with cosmological parameters. However, the amplitude $\mathcal{A}(r) = \mathrm{min } \{|v_u|,|v_l|\}$ of these caustics, defined as being half the difference between the upper and the lower line-of-sight velocities $v_u$ and $v_l$ does provide a measure of the gravitational potential $\phi(r)$ of the group. The method has been subsequently used by \citet{diaferio_caustic_2005, diaferio_measuring_2009, serra_measuring_2011} as a robust way of calculating masses of galaxy groups using all of the positional information from a group (out to $R_{200}$) and as a test for cluster membership \citep{rines_cirs:_2006}. Additional research has shown that for a set of rich X-ray luminous clusters, caustic mass estimates agree to within a ratio of 1.03 $\pm$ 0.11 with mass estimates obtained via lensing analysis \citep{rines_cairns:_2003}.

In this work we apply the caustic method to the Galaxy and Mass Assembly (GAMA\footnote[1]{http://www.gama-survey.org/}) Galaxy Group catalogue (hereafter G$^3$Cv1; \citealp{robotham_galaxy_2011}) and attempt to provide complementary caustic mass estimates to the dynamical mass estimates of the group haloes within the catalogue with the aim of verifying these masses. By also applying the method to mock light cones that mimic the GAMA data we are able to carefully calibrate our algorithm to produce median-unbiased mass estimators for each galaxy group using only redshift and positional information out to radii that are well beyond the virial radius.

The GAMA project \citep{driver_gama:_2009, driver_galaxy_2011} is an ongoing major galaxy survey covering 21 bands of the electromagnetic spectrum from the ultraviolet through to radio wavelengths using a multitude of ground and space based telescopes. The aims of the survey are to cover a region of $\sim 360$ deg$^2$ and to obtain $\sim$ 400,000 redshifts for galaxies to a magnitude of $r_{\mathrm{AB}} = 19.8$ mag. During phase one of GAMA we observed three fields of $4 \times 12$ degrees centred at $\alpha$ = 9h, $\delta$ = 1 deg (G09), $\alpha$ = 12h, $\delta$ = 0 deg (G12) and $\alpha$ = 14.5h, $\delta$ = 0 deg (G15). In phase two we are expanding these to $5 \times 12$ degrees as well as gathering data in three new fields. One of the principal scientific goals of GAMA is to understand and better constrain the halo mass function (HMF; \citealp{Press1974,Lacey1994,Sheth2001,Warren2006,Tinker2008}), which describes the relationship between the mass and number density of dark matter haloes. To that end, it is very important for any catalogue of galaxy groups produced to have accurate and reliable mass estimates, particularly for low-mass groups.

This paper is structured as follows: in Section 2 we briefly describe the theoretical framework behind the caustic mass estimation method, our changes to the algorithm, and a step-by-step description of its implementation. In Section 3 discuss the data set used in this work (mock galaxy group catalogues created for GAMA and the actual groups themselves), and our results are presented in Section 4. We summarise our results in Section 5. Throughout this paper, consistent with the cosmology used to create the GAMA mocks and \citet{robotham_galaxy_2011}, we use a cosmology of $\Omega_{\mathrm{m}} = 0.25,\; \Omega_{\Lambda}=0.75,\; H_0 = h\; 100 \mathrm{km s}^{-1}\;\mathrm{Mpc}^{-1}$.

\section{Methodology}

The caustic method relies on analysing the distribution of group members in a redshift space diagram, which is defined in D99 as the plane $(r, v)$ of the galaxies, where $r$ is the projected radial separation from the group centre, and $v$ is the line-of-sight velocity relative to the group centre of mass. The spherical infall model of \citet{regos_infall_1989} predicts the existence of two trumpet shaped `lines' on this plane where the phase-space density in redshift space is infinite; and in practice these trumpets are observed when looking at both simulated and real groups. By definition, galaxies outside of these caustics are beyond the turnaround radius of the group. If we assume that galaxies lying outside the caustics are considered to be escaping the group, it follows that the caustic describes the escape velocity $v_{\mathrm{esc}}^2 (r)$ of a cluster as a function of distance $r$ from its centre. The term `caustic' refers to the formation of a singularity at a location where the Jacobian of a co-ordinate transformation vanishes; in the case of the caustic method, this is the transformation from real to redshift space. In the spherical model, this causes the galaxies within a group to collapse to the peak redshift of the group along the line-of-sight, placing them somewhere between the centre of the group and the turnaround radius. In redshift space this translates to the trumpet-like lines that describe the escape velocity of the group. 

Here follows a brief review of the physical justification behind the caustic method. Full details of the model can be found in DG97, D99 and \citep{serra_measuring_2011}. Assuming a spherically symmetric model, the escape velocity within a shell of radius $r$ for a group is given by $v_\mathrm{e}^2(r) = -2\phi(r)$. Given our position as observers, it is the line-of-sight component ($v_{\mathrm{los}}$) of this escape velocity that determines the location of the caustics, but this value depends upon the escape velocity profile of the cluster which we may not always know. Instead, we require an expression for the caustic amplitude that is \emph{independent} of the escape velocity profile. \citet{serra_measuring_2011} determine such an expression via the thought process summarised below:

We begin by looking at  the velocity anisotropy parameter $\beta(r) = 1 -( \langle v_{\theta}^2\rangle + \langle v_{\phi}^2 \rangle)/2\langle v_r^2 \rangle$ where $v_{\theta}$, $v_{\phi}$ and $v_r$ correspond to the longitudinal, azimuthal and radial components of an individual galaxy's velocity respectively. If we assume that cluster rotation is negligible, $\langle v_\theta^2 \rangle = \langle v_\phi^2 \rangle = \langle v_{\mathrm{los}}^2 \rangle$ and $\langle v_r^2  \rangle = \langle v^2 \rangle - 2 \langle v^2_{\mathrm{los}} \rangle$ \textbf{where $v_{\mathrm{los}}$ is the line-of-sight component of the velocity}. Rearranging this for $\langle v^2 \rangle$ and incorporating the equality of all velocity components into the expression for $\beta(r)$ gives:

\begin{equation}
	\langle v^2 \rangle = \langle v_{\mathrm{los}}^2 \rangle \left(\frac{3-2\beta(r)}{1-\beta(r)}\right)  \equiv \langle v^2_{\mathrm{los}} \rangle g(r) 
\end{equation}

\noindent where

\begin{equation}
g(r) = \frac{3-2 \beta(r)}{1-\beta(r)} = \frac{2 \langle v_{\mathrm{los}}^2 \rangle + \langle v_r^2 \rangle}{\langle v_{\mathrm{los}}^2 \rangle}
\end{equation}

Given that the potential of a system is related to its escape velocity in the form of $-2 \phi = \langle v_{\mathrm{esc}}^2 (r) \rangle$, it is possible to link the potential to the caustic amplitude if we make the assumption that $\mathcal{A}^2 (r) = \langle v_{\mathrm{esc,los}}^2 \rangle$ in the form 

\begin{equation}
-2 \phi (r) = \langle v_{\mathrm{esc,los}}^2 \rangle g(r) = \mathcal{A}^2(r) g(r)
\end{equation}

\noindent which reduces our problem to having only one unknown: the parameter $\beta$. The final link in the chain is to consider the mass of an infinitesimal shell: 

\begin{equation}
G \ud m = -2 \phi(r) \mathcal{F}(r) \ud r = \mathcal{A}^2 (r) g(\beta) F(r) \ud r
\label{eqn:shell}
\end{equation}

\noindent with $\mathcal{F}(r) = \frac{-2 \pi G \rho (r) r^2}{\phi (r)}$. Combining the expression for $\mathcal{F}(r)$ and $G \ud m$ brings us back to the familiar result for the mass of a shell; $G \ud m = 4 \pi G \rho (r) r^2 \ud r$. Integrating Eq. \ref{eqn:shell} gives

\begin{equation}
GM(<r) = \int_0^r \mathcal{A}^2(r) g(r) F(r) \ud r
\end{equation}

This expression is close to what we need, but is limited by the fact that the density profile of the system needs to be known in order to get a value for the mass. To overcome this, D99 assume that $\mathcal{F}_\beta(r) = \mathcal{F}(r) g(\beta)$ is a slowly varying function with respect to the radius of the system in hierarchical clustering scenarios, and this result is confirmed by \citet{serra_measuring_2011}. We therefore set $\mathcal{F}_\beta (r)$ to be constant and adopt the value of $\mathcal{F}_{\beta} = 0.7$ of \citet{serra_measuring_2011} (this is not critical, as we later adjust our mass estimates by a scaling factor, discussed below), giving the final expression

\begin{equation}
GM(<r) = \mathcal{F}_\beta \int_0^r \mathcal{A}^2 (r) \ud r
\end{equation}

\noindent which can be used to provide mass estimates for our data sample.

\subsection{Caustic mass estimation algorithm}

A successful caustic mass estimation algorithm must be able to successfully infer the continuous dark matter distribution in the halo from a discrete set of points determined by galaxies. The most important goal of our algorithm is to correctly determine the location of the caustic for a group. To do this, we project the group into an area of redshift space and generate a kernel that describes the continuous density of the group within this area \citep{pisani_non-parametric_1993}. Based on this density distribution the algorithm then determines a threshold at which the caustic is placed. 

For a given galaxy group for which member positions are known ($\alpha$, $\delta$, and $z$), the projection into redshift space takes place via the following transformations:

\begin{equation}
r = \frac{c D_A (z_c)}{H_0} \tan \psi
\end{equation}

\noindent and

\begin{equation}
v = c \frac{z-z_c}{1+z_c}
\end{equation}

\noindent where $D_A$ is the comoving distance to the galaxy, $z_c$ is the redshift of the group centre, and $\psi$ is the angular separation of a member galaxy from the group centre at redshift $z$ along the line of sight. The area of redshift space is therefore determined by the full radial extent of the group and its range of line-of-sight velocities. Consider $N$ galaxies in a cluster distributed in a redshift diagram with coordinates $\mathbf{x} = (r,\,v)$. Using an adaptive kernel method \citep{silverman_density_1986}, we describe the density distribution of these galaxies as

\begin{equation}
f_q(\mathbf{x}) = \frac{1}{N} \sum_{i=1}^N \frac{1}{h_i^2} K \left(\frac{\mathbf{x} - \mathbf{x}_i}{h_i}\right)
\label{eqn:density}
\end{equation} 

\noindent where $K$ is the adaptive kernel

\begin{equation}
K(\mathbf{t}) = \left\{
	\begin{array}{ll}
	4 \pi^{-1} \left(1-t^2\right)^3  & \textrm{if } t < 1, \\
	0  & \textrm{otherwise}
	\end{array} \right.
	\label{eqn:kernel}
\end{equation}

\noindent and $h_i = h_c h_{\mathrm{opt}} \lambda_i$ is the local smoothing parameter. $\lambda_i = \left[\gamma / f_1(\mathbf{x}_i)\right]^{1/2}$ where $f_1$ is equation (\ref{eqn:density}) where $h_c = \lambda_i = 1$ for any $i$ and $\log \gamma = \sum_i \log \left[f_1(\mathbf{x}_i)\right]/N$. The motivation for using an adaptive kernel estimator is to have a density estimator that can adapt to density distributions where the true probability density changes quickly \citep{pisani_non-parametric_1993}; this is generally a caveat of fixed kernel estimators that risk to oversmooth or undersmooth the probability distribution. Finally, the optimal smoothing parameter $h_{\mathrm{opt}}$ is

\begin{equation}
h_{\mathrm{opt}} = \frac{3.12}{N^{1/6}} \left(\frac{\sigma_r^2 + \sigma_v^2}{2}\right)^{1/2}
\label{eqn:smoothing}
\end{equation}

\noindent where $\sigma_r$ and $\sigma_v$ are respectively the uncertainties in the galaxy coordinates. The positional uncertainty $\sigma_r$ is calculated from the astrometric uncertainties in GAMA and is negligible, while for $\sigma_v$ we use the same value of $55$ km s$^{-1}$ as used in \citet{robotham_galaxy_2011}.

Performing this calculation can take a great deal of time, particularly when it comes to calculating $f_q$ and optimizing for the best value of $h_c$. A faster, time saving way of obtaining the density estimator is to use Fast Fourier Transforms to convolve a two-dimensional histogram of the data in redshift space with the adaptive kernel \citep{silverman_density_1986}. Other alternative implementations of the algorithm include using friends-of-friends algorithms and binary trees, but in our case, where the galaxy groups have already been selected and statistically well-defined, the FFT approach works best. The process is described in \citet{silverman_density_1986} for one dimension, but can easily be extended to two.

\begin{figure}
	\centering
	\includegraphics[width=0.5\textwidth]{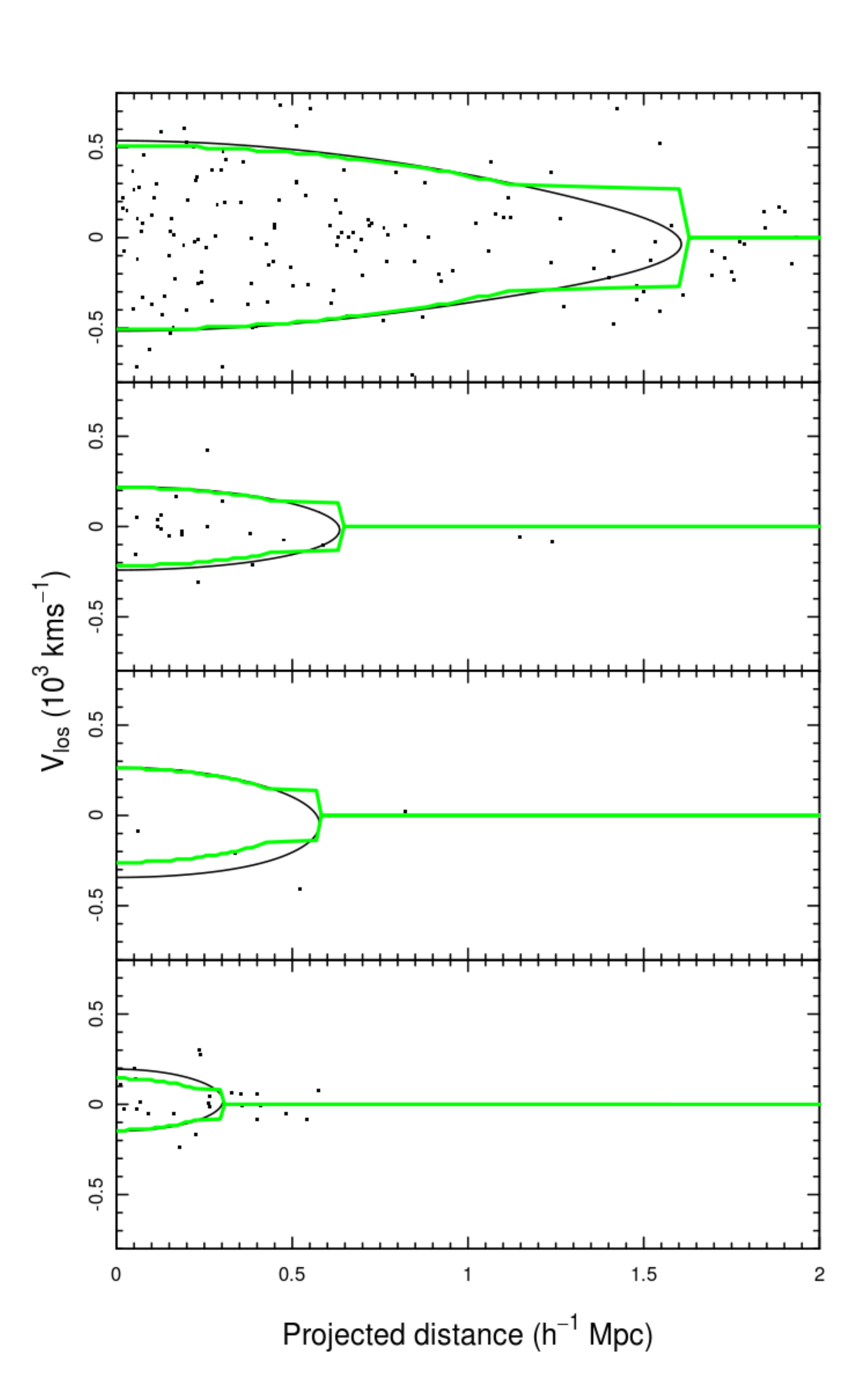}
	\caption{Four examples of the placement of the contour $\kappa$ (black) and the caustics fitted to it (green) for four FoF mock galaxy groups (whose galaxies are shown as the black points) from the G$^3$Cv1 in descending order of luminosity. It is evident here that the FFT method used to estimate the density distribution causes the final caustics to be very smooth with respect to the caustics drawn in DG97 and D99. This introduces a source of error in our caustic mass estimates.}
	\label{fig:trumpetplots}
\end{figure}

We begin by creating a two-dimensional normalised histogram of the galaxies in redshift space (projected radius from the group centre versus line-of-sight velocity), and over the same parameter space, a histogram of calculated values for the kernel given in Eq. (\ref{eqn:kernel}). The smoothing parameters $\lambda_i$, $h_{\mathrm{opt}}$ and $h_c$ are used to adjust the values of the data histogram. The density estimate $f_q(r,\,v)$ is defined as the inverse FFT of the product of the forward FFT of the data and kernel histograms, or in other words:

\begin{equation}
f_q(r,\,v) = \mathcal{F}^{-1} \left( \mathcal{F}(\mathbf{data}) \times \mathcal{F}(\mathbf{kernel}) \right)
\label{eqn:fft}
\end{equation}

\noindent where $\mathcal{F}$ and $\mathcal{F}^{-1}$ denote forward and inverse fast Fourier transforms respectively. The end result of this calculation is a two-dimensional matrix that describes the density distribution of the group galaxies in redshift space; note that we use the \emph{modulus} of this matrix in order to discard any phase shifts caused by the Fourier transforms. The matrix has dimensions of $2^8 \times 2^8$; we find that any size lower than $2^7 \times 2^7$ does not provide our algorithm enough resolution to give reliable results. The caustics are drawn on the density distribution on locations where $f_q(r,\,v) = \kappa$, and $\kappa$ is obtained by minimising the function $S(\kappa,R)$ taken from D99:

\begin{equation}
S(\kappa,R) = | \langle v_{\mathrm{esc}}^2\rangle_{\kappa,R} - 4 \langle v^2 \rangle_R | ^2
\label{eqn:kappa}
\end{equation}

\noindent where the term $\langle v_{\mathrm{esc}}^2\rangle_{\kappa,R}$ is the square of the mean escape velocity at $R$ for a given value of $\kappa$. This corresponds to the average size of the caustic amplitude from the group centre to the maximum projected radial distance $R$ for a given value of $\kappa$ and $\langle v^2 \rangle_R$ is the group velocity dispersion taken from the G$^3$Cv1. To minimise the function $S(\kappa,R)$, we wish to find the value of $\kappa$ for which the average caustic size is equal to $4 \langle v^2 \rangle_R$. We use the R function \textsc{optim} to do this; \textsc{optim} is a general-purpose optimisation function based around the Nelder-Mead algorithm \citep{Nelder1965}, which lends itself particularly well to this task as $S(\kappa,R)$ is a parabolic function with only one well-defined minimum. Once the location of the contour is determined (black line in Figure \ref{fig:trumpetplots}), the algorithm draws the caustics (green lines in Figure \ref{fig:trumpetplots} along $\mathrm{min } \{|v_u|,|v_l|\}$ where $v_u$ and $v_l$ correspond to the upper and lower values of the line-of-sight velocity of the group along the contour. The algorithm scans through the density distribution in bins of $r$, and for each bin selects the minimum of these two velocities ($v_u(r)$ and $v_l(r)$, and reflects it along the line-of- sight velocity axis. The caustic amplitude beyond the maximum extent of the group is artificially set to 0, even though often the caustic closes before the maximum radial extent of the group is reached (see Figure \ref{fig:trumpetplots}).

Based on mock catalogues of galaxies built with N-body simulations in DG97, there is a constraint on the logarithmic derivative of the caustic amplitude: $\ud \ln \mathcal{A} / \ud \ln R \le 2$. Any values of $\mathcal{A}(r)$ for which this derivative does not hold are considered to be the result of the caustic algorithm coming up with the wrong location for the caustic at that particular radius, often due to excessive foreground/background galaxies. Instead, in these cases we use a value for the caustic amplitude such that $\ud \ln \mathcal{A} / \ud \ln R = 1/4$, as in \citet{serra_measuring_2011}. 

To summarise, our algorithm works as follows:  
\begin{enumerate}
\item
Convert the galaxy positions in redshift space into a two-dimensional histogram of galaxy number densities,
\item
create another histogram of the same dimensions containing values for the kernel as per Eq. (\ref{eqn:density}),     
\item 
calculate $f_q(r,\,v)$ using Eq. (\ref{eqn:fft}), 
\item 
calculate the best value for $\kappa$ with Eq. (\ref{eqn:kappa}),
\item
fit the caustics by reading off the minimum value of $|v_u|$ and $|v_l|$ along $r$, whilst ensuring that the derivative inequality holds, 
\item
integrate between the caustics to estimate the mass of the group using Eq. (\ref{eq:finalmass}) and scale it accordingly to obtain a median-unbiased estimator. 
\end{enumerate} 
It is important to stress that this FFT implementation of the kernel estimator still retains the calculations and expressions given in this section; we simply speed up the procedure of calculating the density distribution by applying the kernel with a FFT. 

Our final mass expression is

\begin{equation}
\frac{M_\mathrm{c}}{h^{-1} \mathrm{M}_\odot} = \frac{0.7 A_{\mathrm{c}}}{G / (\mathrm{M}_\odot \mathrm{km}^2\mathrm{s}^{-2} \mathrm{Mpc})} \int_0^r \mathcal{A}(r)^2 \ud r
\label{eq:finalmass}
\end{equation}

\noindent where $r$ is given in units of $h^{-1}$ Mpc and $\mathcal{A}$ in s$^{-1}$ km.  $\int_0^r \mathcal{A}^2(r) \ud r$ is calculated by discretising $\mathcal{A}(r)$ over a set of equally spaced steps and $A_{\mathrm{c}}$ is the caustic mass scaling factor. In Figure \ref{fig:trumpetplots} we show example caustic fits for four friends-of-friends mock galaxy groups of descending total luminosity (from $10^{12}$ h$^{-2}$ L$_{\odot}$ to $10^{9}$ h$^{-2}$ L$_{\odot}$).

Despite the computational efficiency of the FFT method to calculate the density estimate, there are a number of drawbacks that need to be adressed. To begin with, the area over which the 2D histograms for the data and the kernel are created need to be larger than the area the data spans. This is to avoid the kernel (and the resulting density estimate) wrapping around the borders due to the periodic nature of Fourier transforms and results in a very smooth density distribution compared to those shown in DG97 and D99. This will effectively artificially increase the sizes of our caustics and cause a systematic overestimation of the caustic mass (though the scaling factor $A_c$ corrects for this); the smoothing is independent of group size as shown in Figure \ref{fig:trumpetplots}. We do not consider the presence of background galaxies in this analysis, and make the assumption that the friends-of-friends algorithm used in \citet{robotham_galaxy_2011} has recovered group members as accurately as possible. This assumption is tested in Section 3.

\section{Data}

\subsection{The GAMA Galaxy Group Catalogue}

\begin{figure*}
	\centering
	\includegraphics[width=1.0\textwidth]{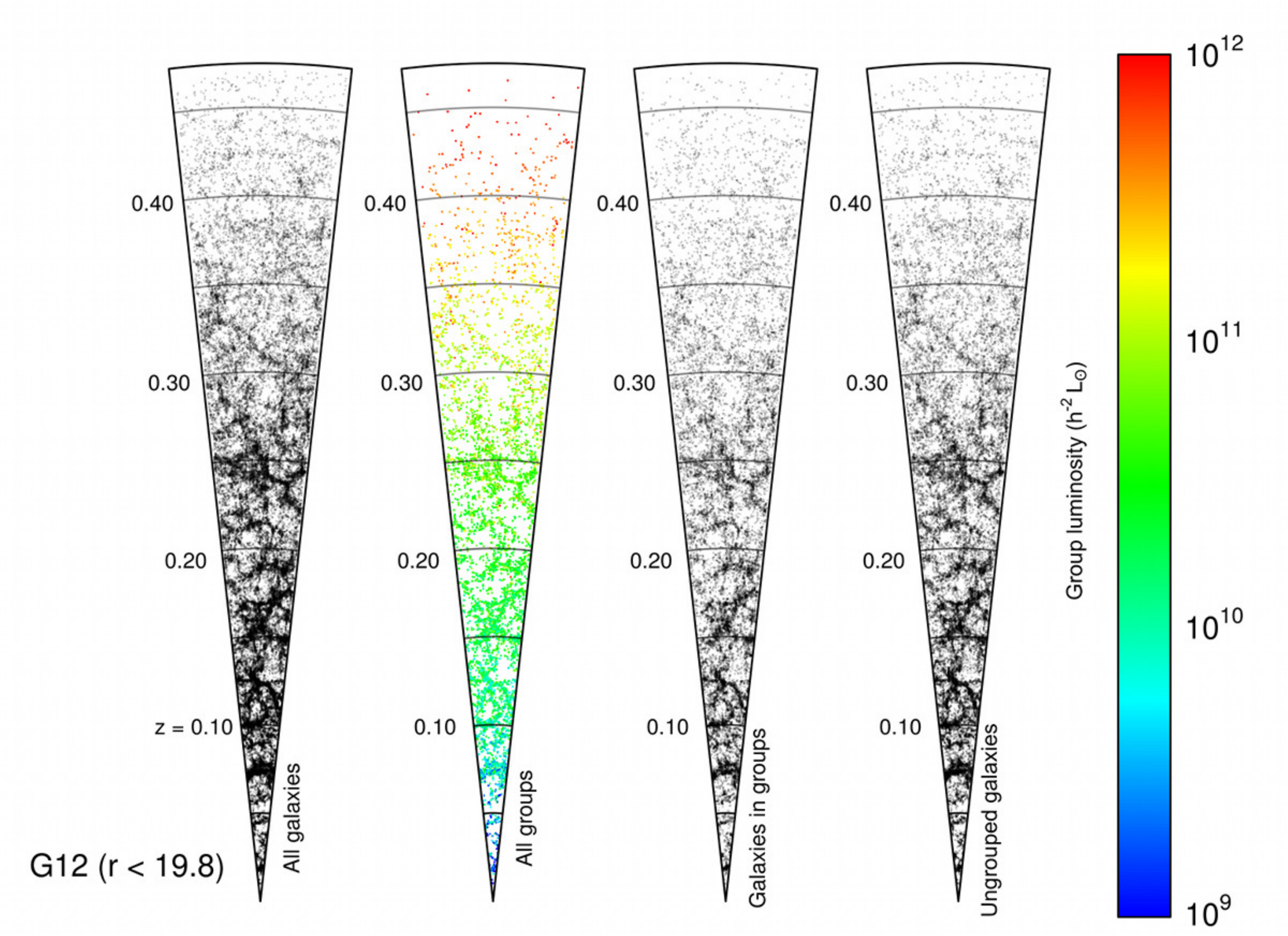}
	\caption{Example of a cut-out cone showing galaxies and groups with $r_{\mathrm{AB}} < 19.8$ for the GAMA G12 field going from $174^\circ \leq \mathrm{RA} \leq 186^\circ$ out to redshift $z \sim 0.5$. From left to right, the cones display all galaxies in this field, all groups found in G$^3$Cv1 (coloured by their total group luminosity), all grouped galaxies and all ungrouped ones.}
	\label{fig:cones}
\end{figure*}

The groups presented in the G$^3$Cv1 have been identified using a slightly modified friends-of-friends algorithm \citep{press_how_1982,huchra_groups_1982} to correctly account for the distortions due to peculiar velocities. It achieves this by considering projected separations independently of radial positions. Figure \ref{fig:cones} shows cone diagrams containing galaxies and groups for the G12 field, where it is possible to see how they trace the large-scale-structure.

The final group catalogue contains approximately 14,000 galaxy groups with $\geq 2$ members out to $r_{\mathrm{AB}} = 19.4$ (fields G09 and G15) and 19.8 mag (G12 field) which encompass about 40\% of the galaxies in the full GAMA catalogue. Robotham et al. estimated their masses by using the relation $M \propto \sigma^2 R$ where $\sigma$ and $R$ correspond to the velocity dispersion and the projected group radius. The group velocity dispersion is calculated using the \textsc{gapper} algorithm given in \citet{beers_measures_1990} where all the recession velocities for $N$ galaxies within a group are ordered, and the velocity dispersion is estimated using calculated values for the gaps between velocity pairs $v_{i+1} - v_i$ for $i = 1,\;2...,\;N-1$. Group radius estimates were calculated for radii that contain 50\%, 68\% and 100\% of the galaxies within each group (Rad$_{50}$, Rad$_{68}$ and Rad$_{100}$ respectively), and the projected group centre was located by three different methods, discussed below. The mass of each group was estimated by

\begin{equation}
\frac{M_{\mathrm{FoF}}}{h^{-1} \mathrm{M}_\odot} = \frac{A}{G / (\mathrm{M}_\odot \mathrm{km}^2\mathrm{s}^{-2} \mathrm{Mpc})} \left(\frac{\sigma^2_{\mathrm{FoF}}}{\mathrm{km}^2 \mathrm{s}^{-2}}\right) \frac{\mathrm{Rad}_{\mathrm{FoF}}}{h^{-1} \mathrm{Mpc}}
\end{equation}

\noindent where $G$ is the gravitational constant (6.673 $\times 10^{-20}$ km$^2$ s$^{-2}$ Mpc), $\mathrm{Rad}_{\mathrm{FoF}}$ and $\sigma_{\mathrm{FoF}}$ are the radius and velocity dispersion of the group obtained by the methods described above. $A$ is a scaling factor that is required to obtain a median-unbiased estimate of the friends-of-friends mass with respect to the real halo mass. It varies depending on multiplicity and median redshift of a group. In this work we mimic the multiplicity and redshift subsets of \citet{robotham_galaxy_2011} to ensure a direct comparison. We consider every galaxy in a group when running our algorithm, while the dynamical mass estimates do not contain the full radial galaxy density profile information. The projected group centre is defined in three different ways in the G$^3$Cv1: the first centre corresponds to the $r_{\mathrm{AB}}$ luminosity centre of light (CoL) of all the galaxies associated with the group. The iterative group centre is estimated by calculating the $r_{\mathrm{AB}}$ CoL of the cluster, and then rejecting the most distant galaxy from this centre. This process is then repeated until only two galaxies remain, at which point the brightest one of these is selected as the group centre. In the final method, the brightest group galaxy in the $r_{\mathrm{AB}}$-band is selected as the group centre (BGG). In Section 3 we test the sensitivity of our caustic algorithm against the three group centre definitions.

\subsection{The mock group catalogue}

The mock catalogues were constructed by first populating the dark matter halos of the \emph{Millennium Simulation} \citep{springel_simulations_2005} with galaxies, the positions and properties of which were predicted by the \cite{Bower2006} description of the Durham semi-analytical model, \verb=GALFORM=, and adjusted to match the GAMA survey luminosity function of \citet{Loveday2011}. To generate an all-sky survey to the GAMA depth, it is necessary to stack $11^3$ replicated copies of the simulation box to create a `super-cube' (with sides of length $5500\,{\rm h^{-1}Mpc}$).

Following the prescription of Merson et al. (in prep) to place the halos and galaxies in the super-cube at the position at which they enter the light cone, galaxies are assigned their intrinsic properties, such as stellar mass and fluxes in the appropriate bands. Galaxy properties are generally not interpolated between snapshots, with the exception of fluxes. This is to avoid discontinuities at the snapshot boundaries. Having assigned the r-band apparent magnitudes, they are adjusted to perfectly match the redshift dependent GAMA galaxy luminosity function, following a simple abundance matching in the r-band. Finally the flux limits of the GAMA survey are applied. We then take the all-sky mock catalogue and apply solid angle cuts to create the 9 mocks. 

\section{Caustic mass estimates}

Before applying this algorithm to the actual group catalogue, it is important to understand how well it performs when estimating masses for a set of mock catalogues that have been prepared alongside the G$^3$Cv1. We therefore calibrate our caustic mass estimation algorithm using a set of 9 GAMA mock galaxy catalogues (described in section 2).

In these mock catalogues, the true grouping of galaxies is known, so a well-informed calculation of their halo mass is possible. This acts as a benchmark for our caustic mass estimation algorithm and allow us to experiment with different implementations of the method. \citet{serra_measuring_2011} show that fine-tuning the parameters of the caustic mass algorithm ($h_c,\; q,\;\kappa$) does not provide a considerable improvement of its results. Instead we run the algorithm using different values for the group centre given in the G$^3$Cv1: the centre of light, the iterative group centre and the brightest cluster galaxy. This provides a useful way of testing the stability of the caustic algorithm to different definitions of the group centre, as well as confirming the conclusions from \citet{robotham_galaxy_2011} of which definition is the most appropriate. 

\begin{figure*}
	\centering
	\includegraphics[width=1\textwidth]{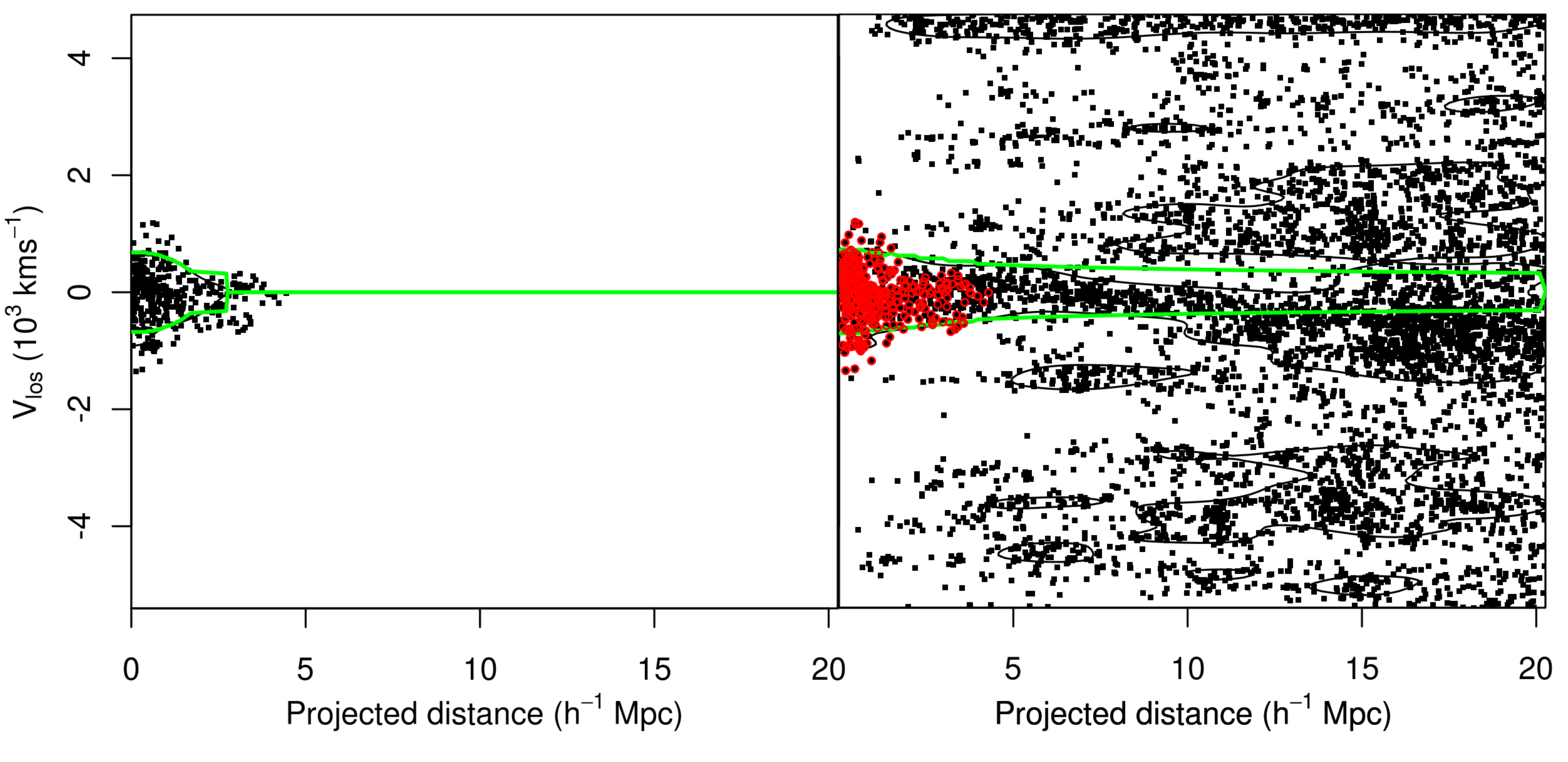}
	\caption{Side-by-side comparison showing how the caustic mass estimation performs for a large mock galaxy group. In both panels, the black points represent the locations of galaxies in redshift space, the black lines show the contour $f_q(r,\,v) = \kappa$ and the green lines are the caustics drawn along $\mathrm{min } \{|v_u|,|v_l|\}$. The left panel is a caustic fit only for galaxies present within this mock group and shows a clear example of the `trumpet' distribution. The right panel includes nearby galaxies in redshift space out to $\pm 4R_{100}$ and $|z-z_{\mathrm{med}}| \leq  4 \times \mathrm{max} |z - z_{\mathrm{med}}|$ whilst circling the original group members in red. In this case the trumpet distribution is lost and the caustics are artificially closed at the end of the sample.}
	\label{fig:HIDcomp}
\end{figure*}

\subsection{Sensitivity to definition of group centre}

In \citet{robotham_galaxy_2011}, the iterative method seems to be particularly robust at picking out the group centre even in the presence of outliers, which tend to throw off the estimated group centre when using the centre of light method. The brightest cluster galaxy approach is also robust to outliers, but analysis reveals that the iterative method recovers more group centres that match the mock groups. It precisely matches the group centre around 90\% of the time, while the centre of light method is the poorest performer because it is the method that depends most explicitly on the group being recovered completely. For the caustic mass algorithm, we expect the most robust definition of the group centre to produce the most stable results; in other words, the variance in the scale factors $A_c$ for each multiplicity and redshift bin should be minimal for the most stable group centre. This is confirmed when we run the caustic algorithm on the mock group catalogue and change only the location of the group centre from the iterative centre-of-light group centre to the brightest group galaxy and then to the centre-of-light: $\sigma^2(A_{\mathrm{IterCen}}) = 0.192$ whereas $\sigma^2(A_{\mathrm{BGG}}) = 0.201$ and $\sigma^2(A_{\mathrm{CoL}}) = 0.361$. We expect the tendency of the centre-of-light method to incorrectly define the group centre to be an outlying bright galaxy to throw off the placement of the caustics by deforming the density distribution of galaxies in redshift space. However, we also expect the caustic algorithm to be robust to minor perturbations in the placement of the galaxies in redshift space, as the difference in the variances given above is minor.

\subsection{Sensitivity to definition of group extent}

Our second exploration involves changing the number of galaxies that we consider when calculating the mass for each group, i.e. artificially increasing group membership. We do this by extending the boundary of the group in redshift space to include some nearby galaxies. In redshift space, this extension is defined as:

\begin{equation}
	|z-z_{\mathrm{med}}| \leq \Delta z; \;\Delta z = 4 \times \mathrm{max} { |z - z_{\mathrm{med}}| }
\end{equation}

\begin{figure}
	\centering
	\includegraphics[width=0.45\textwidth]{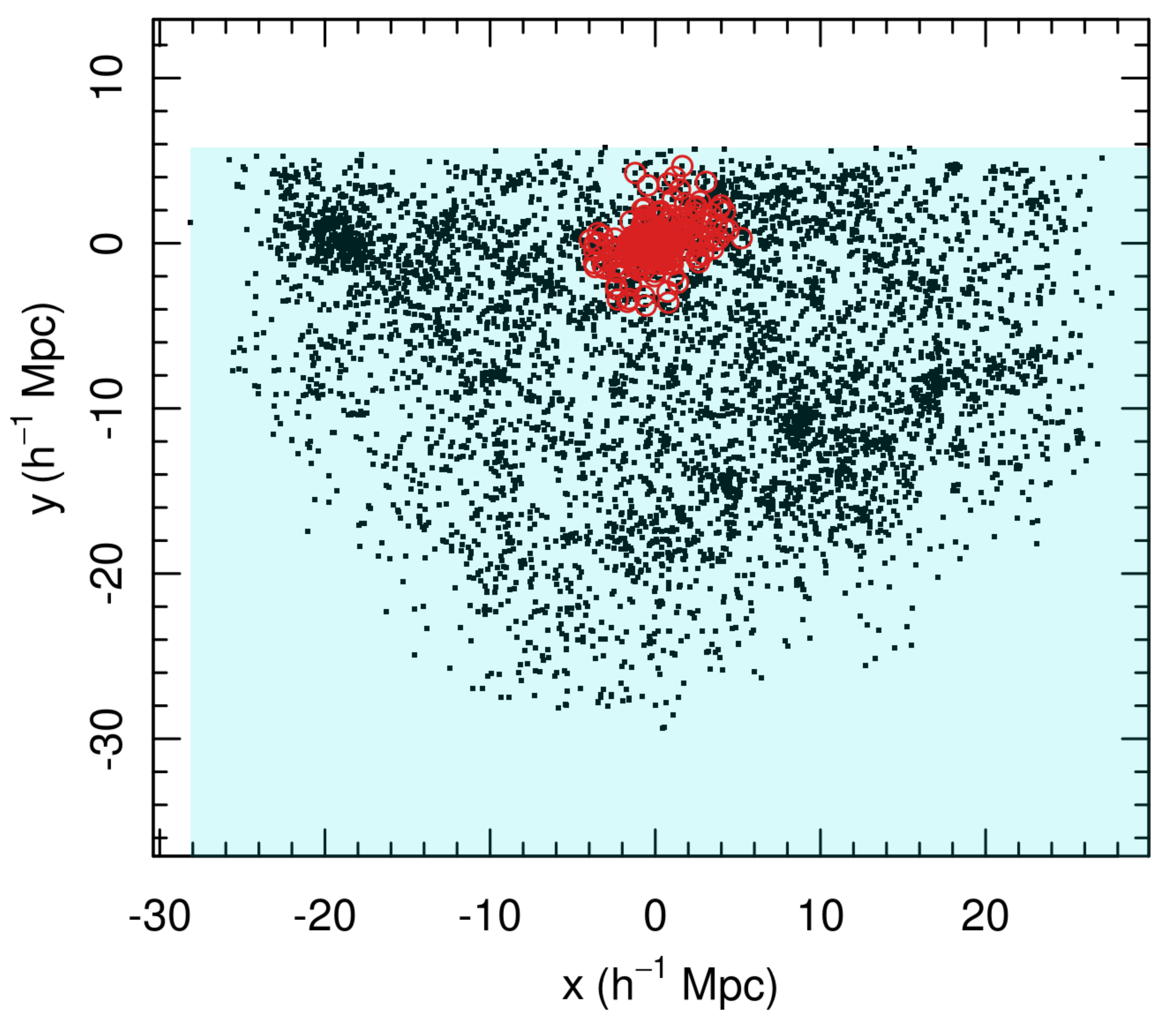}
	\caption{A projection of the mock group in Figure \ref{fig:HIDcomp} showing all galaxies within the mock group circled in red, and all galaxies that are detected when the algorithm includes nearby galaxies out to $\pm 4R_{100}$ and $|z-z_{\mathrm{med}}| \leq  4 \times \mathrm{max} |z - z_{\mathrm{med}}|$ The area shaded in blue represents the physical size of the simulation region at the median redshift of the group. At this range it is evident that the algorithm is including galaxies that are likely to belong to other separate groups.}
	\label{fig:spread}
\end{figure}

Spatially, we increase the maximum distance a given galaxy can be from the group centre. This effectively allows us to check whether the caustic algorithm is sensitive to other interloping groups. We find that including extra galaxies that in some cases belong to other groups not only systematically increases the mass estimates made by the caustic algorithm as one might expect, it also increases the mean spread of the results (defined as the ratio of the logarithm of the true and estimated mass) from $\langle \sigma^2 \rangle = 8.33 \times 10^{-3}$ for caustic mass estimates made with the groups as they are, to $\langle \sigma^2 \rangle = 0.0126$. One expects the overall mass estimate to increase as a result of including more galaxies, but the increase in spread is unexpected: the caustic algorithm ought to work better with a greater number of galaxies. Instead, our algorithm is unable to correctly place the caustics in the redshift space diagram because our extended search cut includes galaxies that are most likely inside other independent groups (Figure \ref{fig:spread}). 

\begin{figure}
	\centering
	\includegraphics[width=0.45\textwidth]{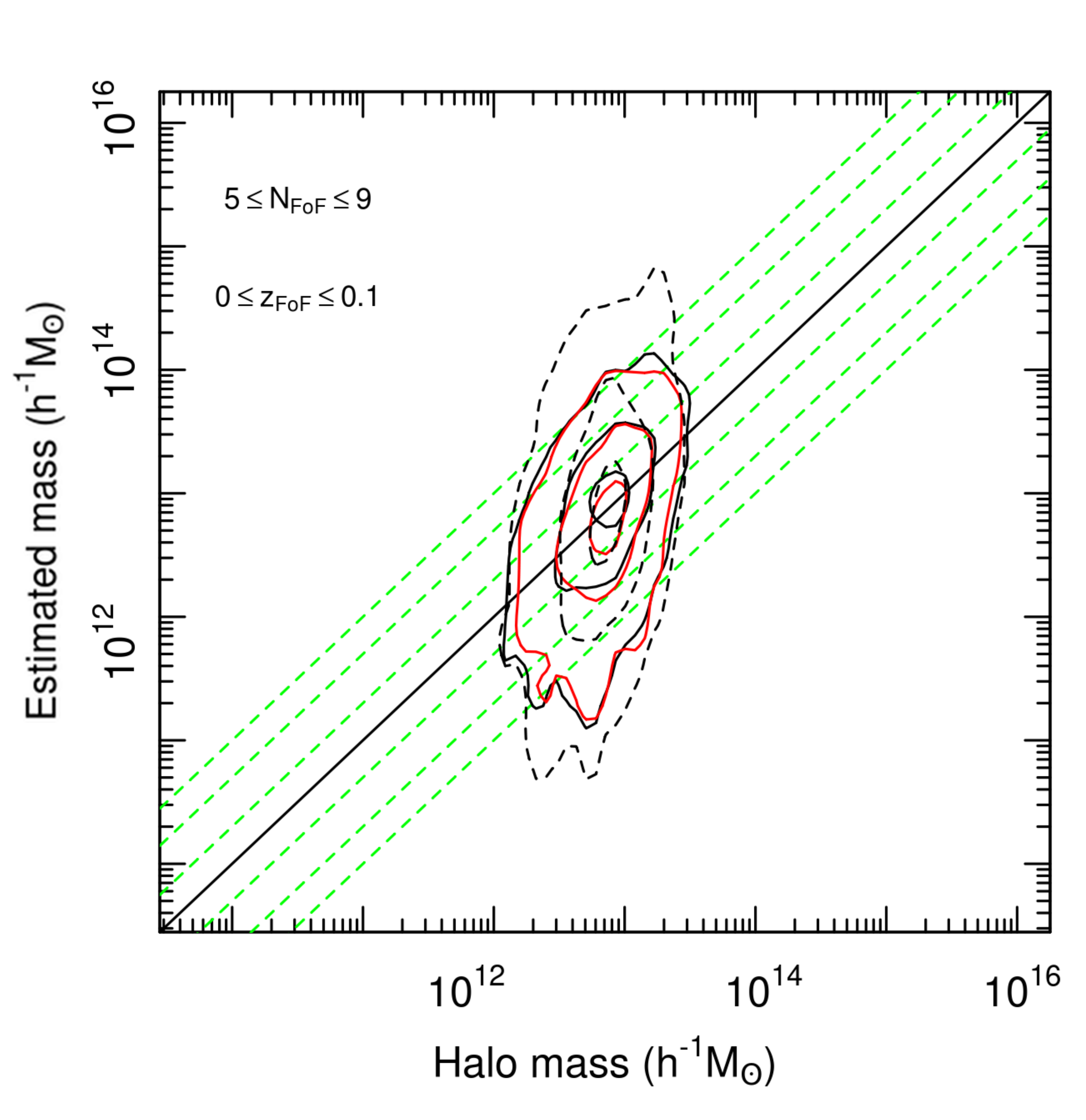}
	\caption{The black dashed contours show the results of running the caustic algorithm on the mock catalogues and considering a larger number of galaxies around each cluster in the catalogues. By comparison, the solid black contours represent the mass estimates made when considering only galaxies known to be in each group from the simulation. Finally, the red contours represent the distribution of the dynamical mass estimates from the G$^3$Cv1. All three distributions have been adjusted to the same median as the dynamical mass estimates, shown in the red contours. Each contour line contains 10, 50 and 90\% of the points, and the dashed green lines are 2/5/10 times away from the median.}
	\label{fig:bigmultex}
\end{figure}

Visual examination of the redshift space plots in Figure \ref{fig:HIDcomp} of these extended group cuts demonstrates that the inclusion of nearby galaxies disrupts the distinctive trumpet shaped distribution seen in ideal spherical groups. Figure \ref{fig:bigmultex} shows a mass comparison between the group catalogue data and the extended group cut. Shown are a subset of groups; those that with redshift between 0 and 0.1, and with 5 to 9 members (before including nearby galaxies). The caustic mass estimates made with the extended group cut (in dashed black lines) show a much greater spread, due to the reasons described above. 

This result highlights the importance of carefully determining group membership when using the caustic algorithm, as the presence of galaxies that are not associated with the group being considered can have a catastrophic effect on the locations of the caustics, ultimately resulting in an incorrect mass estimate. By combining the caustic method with the friends-of-friends algorithm used to determine group membership for the G$^3$Cv1 and mocks we are able to significantly reduce the probability of the caustics being placed incorrectly due to the presence of interloper galaxies within a group.

\subsection{Application to mocks}

\begin{figure*}
	\centering
	\includegraphics[width=1\textwidth]{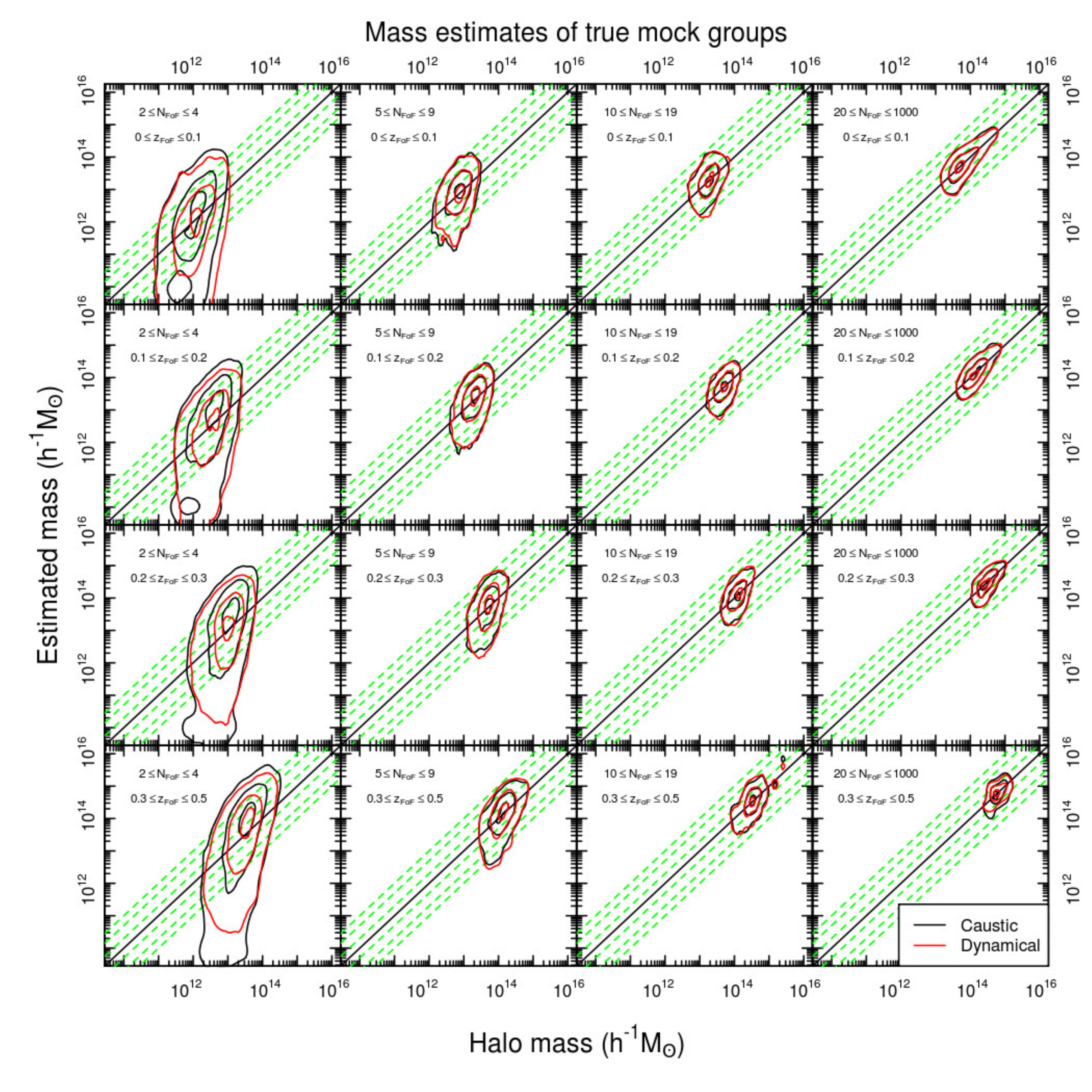}
	\caption{Distribution of caustic (dynamical) masses for the intrinsic mock groups as a function of halo mass, drawn in black (red) for $r_{\mathrm{AB}} \leq 19.4$. For each panel, both mass estimates have been corrected to be median-unbiased. The contours represent areas containing 10, 50 and 90\% of the groups and the green lines are regions where the mass estimate is 2/5/10 times off the true mass. Of particular interest is the tendency for both distributions to follow each other very closely, particularly when over or under estimating the true halo mass.}
	\label{fig:mockscat_194}
\end{figure*}

Using the original grouping from the G$^3$Cv1 with the iterative group centre, we calculate caustic mass estimates for every group in the mock catalogues and calculate $A_c$ for a set of redshift and multiplicity bins in order to make these estimates median unbiased (Table 1). Figure \ref{fig:mockscat_194} shows the results of this process for each multiplicity and redshift bin, comparing the distribution of caustic mass estimates (in black) to the dynamical mass estimates (red). Despite the caustic method performing best in very populated groups ($N \geq 200$; of which there is only one such group in the G$^3$Cv1), demonstrated by the fact that the scatter of the distribution of mass estimates shown in Figure \ref{fig:mockscat_194} decreases as a function of increasing multiplicity. We would normally expect the scaling factor $A_{\mathrm{c}}$ to always equal one. Instead we find that the caustic mass is systematically greater than the masses of the mock groups, i.e. $A_{\mathrm{c}} < 1$. The only exception to this is for groups with two to four members, where the caustic algorithm is much more likely to fail to find appropriate contours, and thus defaults to a specific value, artificially adding a tail to the distribution of masses for that group subset. It is likely that due to the simulated galaxy groups in the mock catalogues not being perfectly spherical, the groups not being virialised (one of the basic assumptions of the spherical infall theory, on which the caustic algorithm is based, is that the group is virialised) the extra smoothing in our caustic introduced by using FFT to calculate the density distribution $f_q(r,v)$, as well as the fact that the caustic method works best for galaxies with more than 200 members, $A_c$ is not always equal to one as it should be in an ideal case. The overall variation in the scale factor $A_c$ is roughly a factor of 4.5, which is of the order of the range of scaling factors used to calibrate the G$^3$Cv1 dynamical mass estimates (see Table 3 in \citet{robotham_galaxy_2011}). In contrast to the dynamical mass estimate scaling factors, the caustic mass scaling factors vary less as a function of redshift, but are far susceptible to variations in group multiplicity.

\begin{table*}
\begin{tabular}{llllllllllll}
 &\multicolumn{2}{l}{$2 \leq N_{\mathrm{FoF}} \leq 4$}& &\multicolumn{2}{l}{$5 \leq N_{\mathrm{FoF}} \leq 9$}& &\multicolumn{2}{l}{$10 \leq N_{\mathrm{FoF}} \leq 19$} & &\multicolumn{2}{l}{$20 \leq N_{\mathrm{FoF}} \leq 1000$}\\
&19.4&19.8& &19.4&19.8& &19.4&19.8& &19.4&19.8\\
\hline
$0 \leq z_{\mathrm{FoF}} \leq 0.1$&1.63&1.63& &0.43&0.43& &0.45&0.46& &0.41&0.41\\
$0.1 \leq z_{\mathrm{FoF}} \leq 0.2$&1.58&1.59& &0.43&0.43& &0.42&0.42& &0.38&0.39\\
$0.2 \leq z_{\mathrm{FoF}} \leq 0.3$&1.52&1.53& &0.42&0.44& &0.36&0.38& &0.35&0.35\\	
$0.3\leq z_{\mathrm{FoF}} \leq 0.5$&1.18&1.21& &0.29&0.31& &0.29&0.29& &0.24&0.26\\
\end{tabular}
\caption{Values for $A_{\mathrm{c}}$ for each group subset in both the $r_{\mathrm{AB}} < 19.4$ and $r_{\mathrm{AB}} < 19.8$ mock G$^3$Cv1 mock catalogues using iterative group centres and with all galaxies in group. Including these numbers in Eq. \ref{eq:finalmass} gives a median-unbiased estimate for the group mass.}
\label{table:scalefactor}
\end{table*}

As expected, the least satisfactory results are for groups that are galaxy pairs, where there is little velocity information for the caustic algorithm to use. This explains the presence of large tails in the $2 \leq N_{\mathrm{FoF}} \leq 4$ panels of Figure \ref{fig:mockscat_194}. However, we must note that both methods tend to fail at these low multiplicities. For higher multiplicity and redshift cuts the caustic mass estimate PDFs shown in this figure agree extremely well with those of the dynamical mass estimates; indicating some correlation between both algorithms when it comes to performing badly for certain galaxy groups. This can be seen in the scatter plots as both contours tend to follow roughly the same profile, meaning that it is likely that a group that performs badly in one algorithm is likely to do so with the other. This is particularly visible for the $10 \leq N_{\mathrm{FoF}} \leq 19,\; 0.3 \leq z_{\mathrm{FoF}} \leq 0.5$ bin where there is a secondary concentration of high mass groups that is present in both distributions.

\begin{figure*}
	\centering
	\includegraphics[width=1\textwidth]{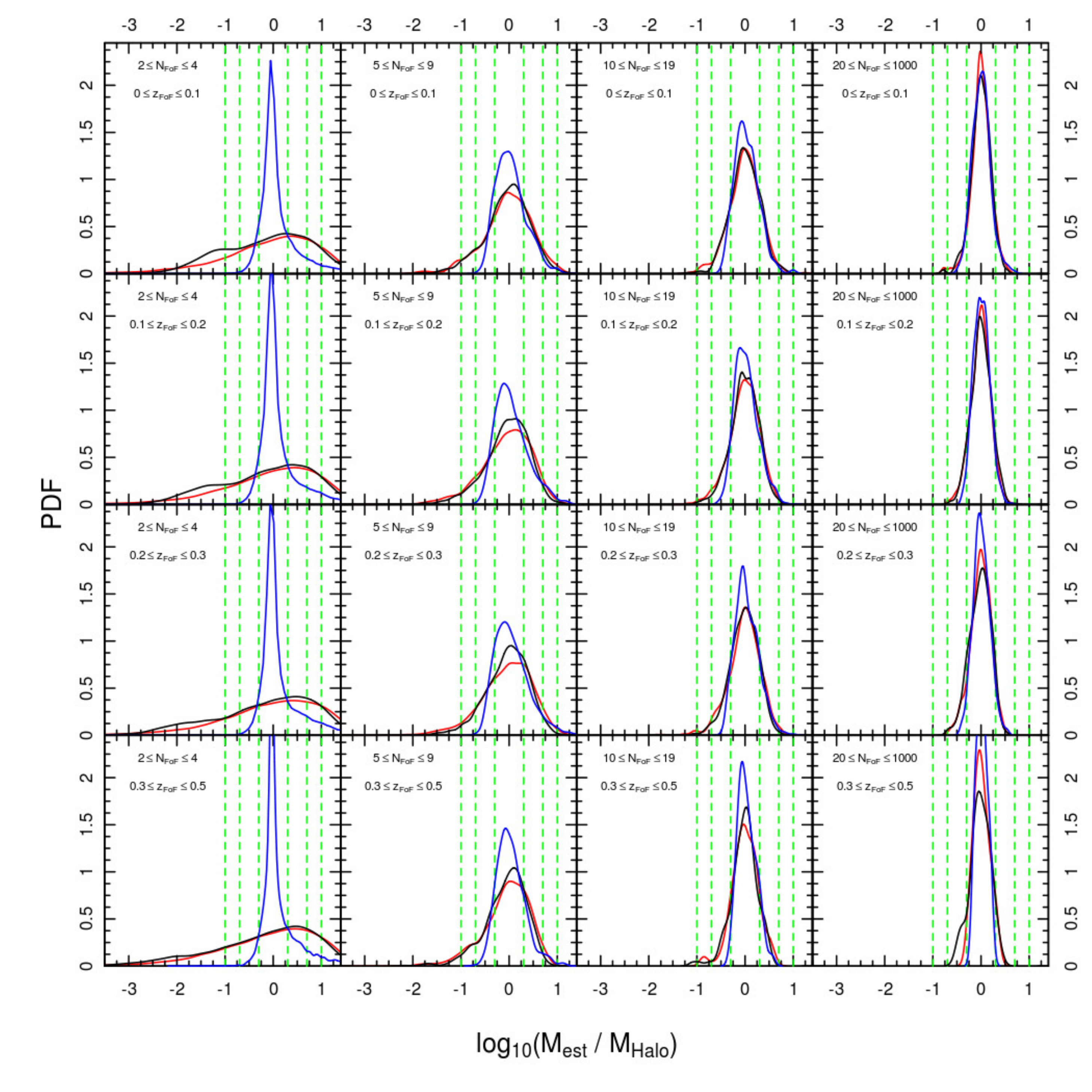}
	\caption{Probability distribution functions of median-unbiased values of $\log \frac{M_{\mathrm{est}}}{M_{\mathrm{Halo}}}$ for the $r_{\mathrm{AB}} < 19.4$ sample, with caustic masses drawn in black, and dynamical masses in red. The blue line is the PDF of $\log \frac{M_{\mathrm{c}}}{M_{\mathrm{dyn}}}$ and highlights the agreement between the two methods. The green dashed lines indicate regions that are factors of 2/5/10 away from the `true' mass. The difference between scatter in the caustic and dynamical mass estimates is often minimal, with the caustic method showing less scatter for groups with $5 \leq N_{\mathrm{FoF}} \leq 19$.}
	\label{fig:mockdens_194}
\end{figure*}

Figure \ref{fig:mockdens_194} translates the information shown in Figure \ref{fig:mockscat_194} into a set of density distributions where the ratio between the caustic and dynamical mass estimates and the known halo mass (shown in black and red respectively) is displayed alongside the ratio between both mass estimates (in blue). As seen on Figure \ref{fig:mockdens_194}, for groups with a mid-range multiplicity ($5 \leq N_{\mathrm{FoF}} \leq 19$) the caustic mass estimates have a greater spread than the dynamical mass estimates; this is true across all redshift bins. For groups with $N_{\mathrm{FoF}} \leq 4$ the scatter is comparable across all redshifts. Both methods produce estimates that are within a factor of 2 in agreement with each other. The large tails seen in Figure \ref{fig:mockscat_194} are visible here: the small `bumps' forming on the right hand of the distribution are seen in both distributions, once again demonstrating that both methods tend to fail in similar ways. Despite the caustic algorithm being designed for high multiplicity groups, we are still able to make reasonable estimates of the group mass.

\begin{figure*}
	\centering
	\includegraphics[width=1\textwidth]{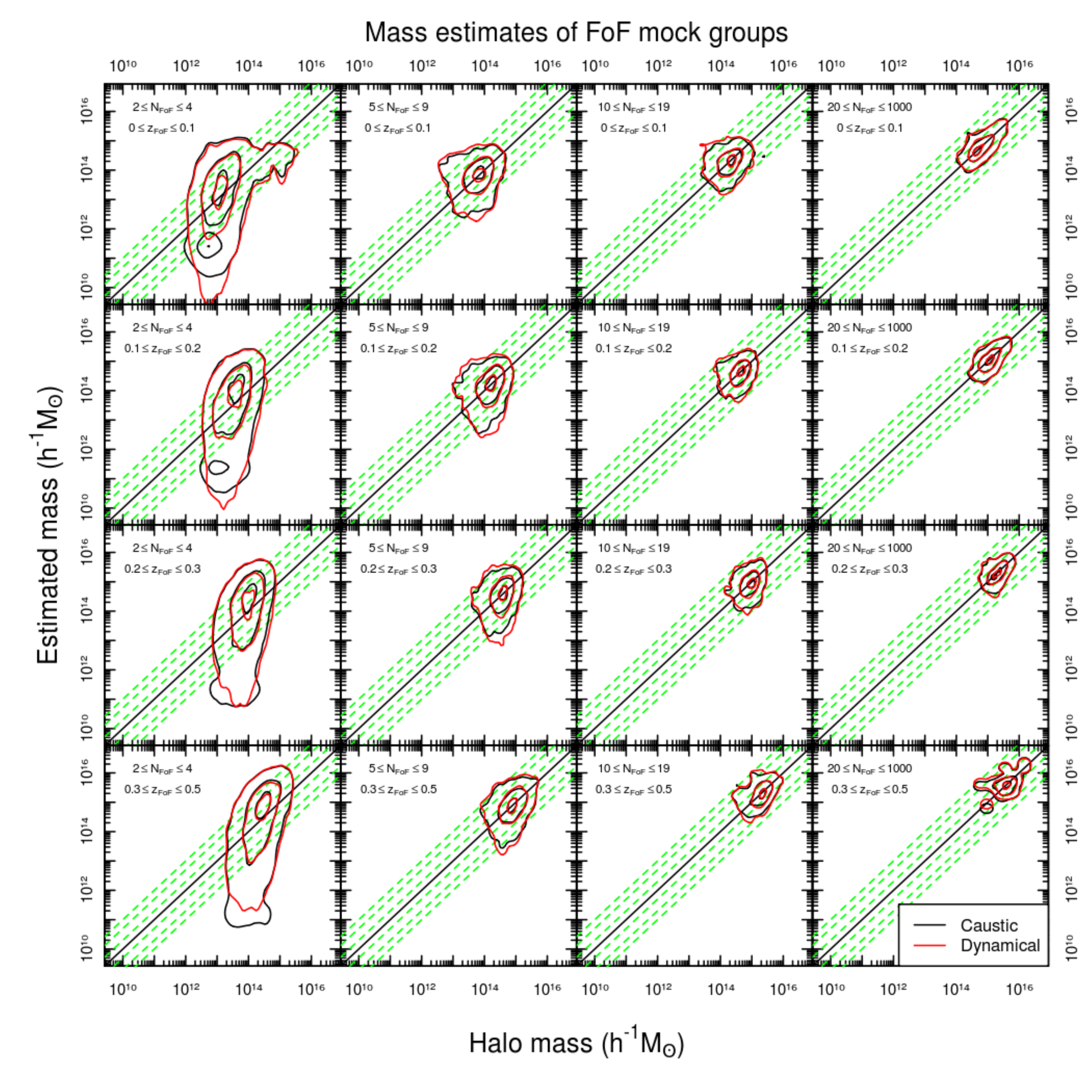}
	\caption{As Fig. \ref{fig:mockscat_194} but for bijectively matched groups identified by the friends-of-friends algorithm instead of the true known intrinsic grouping. The coloured lines represent regions containing 10, 50 and 90 \% of groups. The black contours compare the caustic mass to the FoF halo mass, and FoF dynamical mass estimate to the FoF halo mass estimate are shown by the red contours.}
	\label{fig:mockFoFscat_194}
\end{figure*}

\begin{figure*}
	\centering
	\includegraphics[width=1\textwidth]{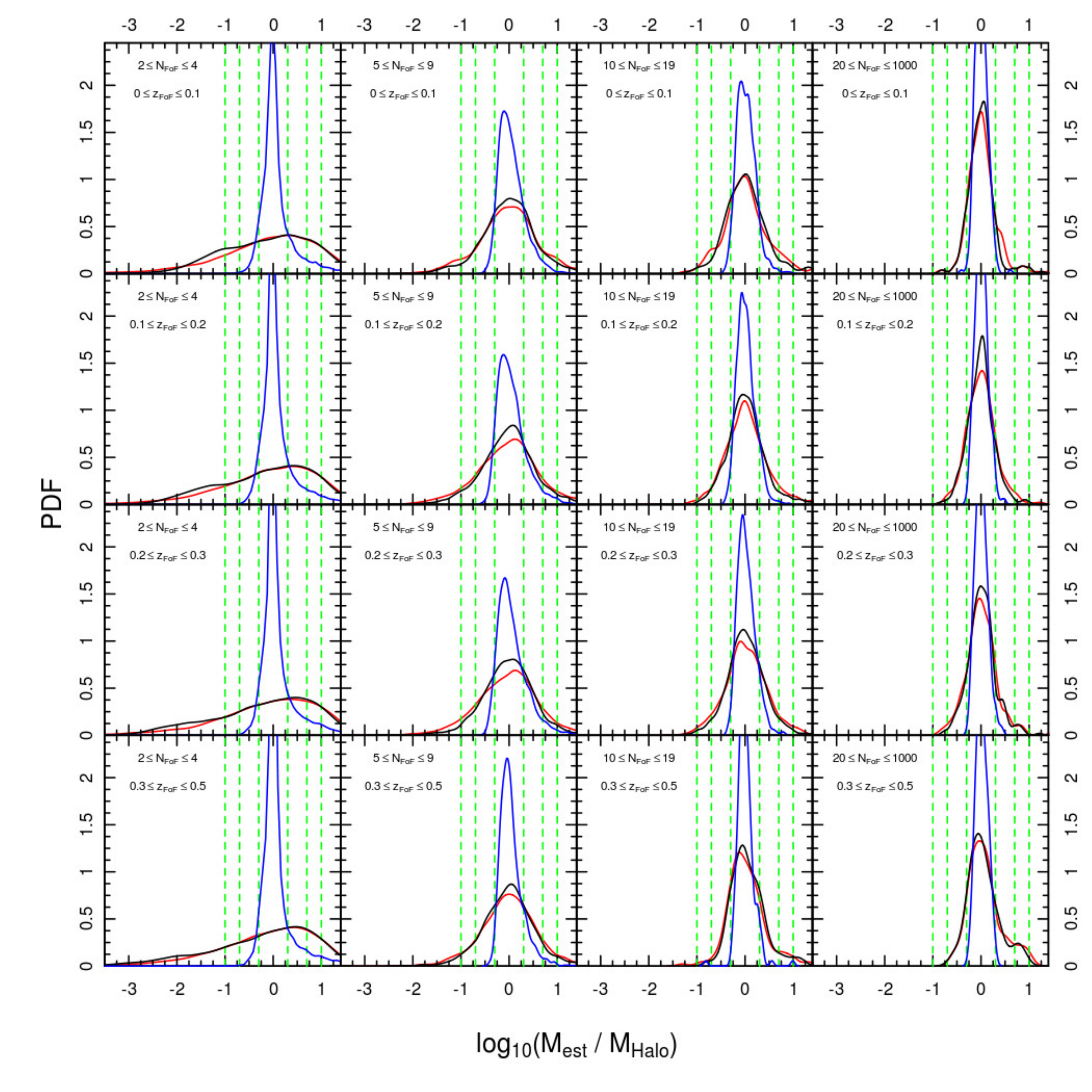}
	\caption{As Fig. \ref{fig:mockdens_194} but for the bijectively matched FoF groups from the mock galaxy catalogues. The different lines show median-unbiased values of $\log \frac{M_{\mathrm{est}}}{M_{\mathrm{Halo}}}$ for the $r_{\mathrm{AB}} < 19.4$ sample. The black line shows the ratio between the caustic mass and the FoF halo mass, while the red line shows the ratio between the FoF dynamical and halo masses. As before, the blue line shows the distribution of the ratio between the caustic mass and the FoF dynamical mass.}
	\label{fig:mockFoFdens_194}
\end{figure*}

Our aim is for the caustic mass estimate to recover the intrinsic halo mass of each group as accurately as possible. However, in practice and when applying the algorithm to the real data in the G$^3$Cv1, we must run the algorithm not on intrinsic groups, but on groups defined by the friends-of-friends algorithm used in \citet{robotham_galaxy_2011}. An important test therefore is to see how the caustic mass estimation algorithm performs on bijectively matched groups drawn out from the mock catalogue using the same friends-of-friends algorithm, and the results of this can be seen in Figures \ref{fig:mockFoFscat_194} and \ref{fig:mockFoFdens_194}. In both figures the distribution of data mimics that of the distributions for the caustic masses on the intrinsic mock groups, with these showing less scatter compared to the caustic masses of the FoF mock groups. Given that the design of the friends-of-friends algorithm used in \citet{robotham_galaxy_2011} to construct the groups in both the mocks and the final group catalogue is to reject groups that have significant outliers, this greater agreement indicates that the caustic mass estimate may be slightly more sensitive to outliers.

\begin{figure*}
	\centering
	\includegraphics[width=0.45\textwidth]{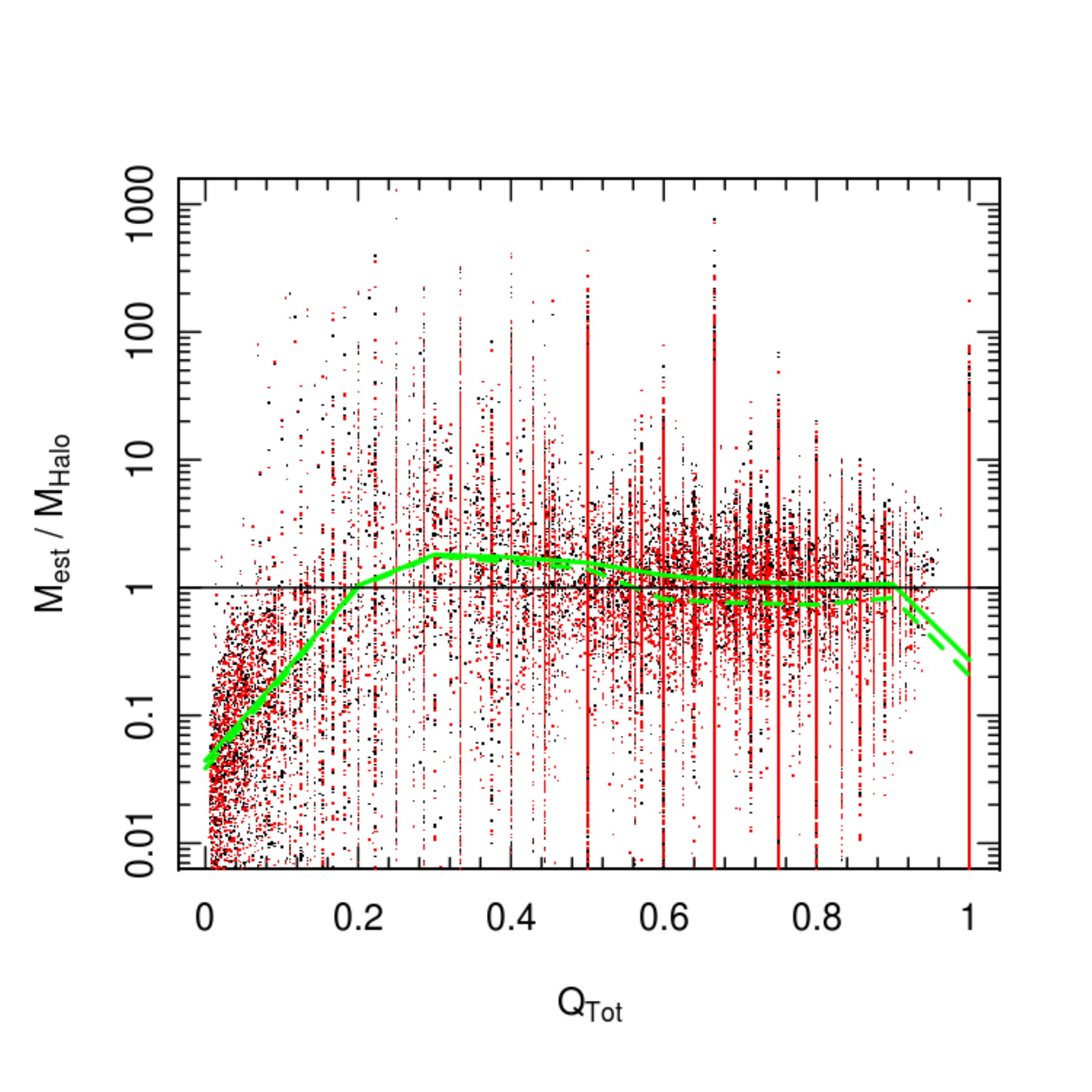}	
	\includegraphics[width=0.45\textwidth]{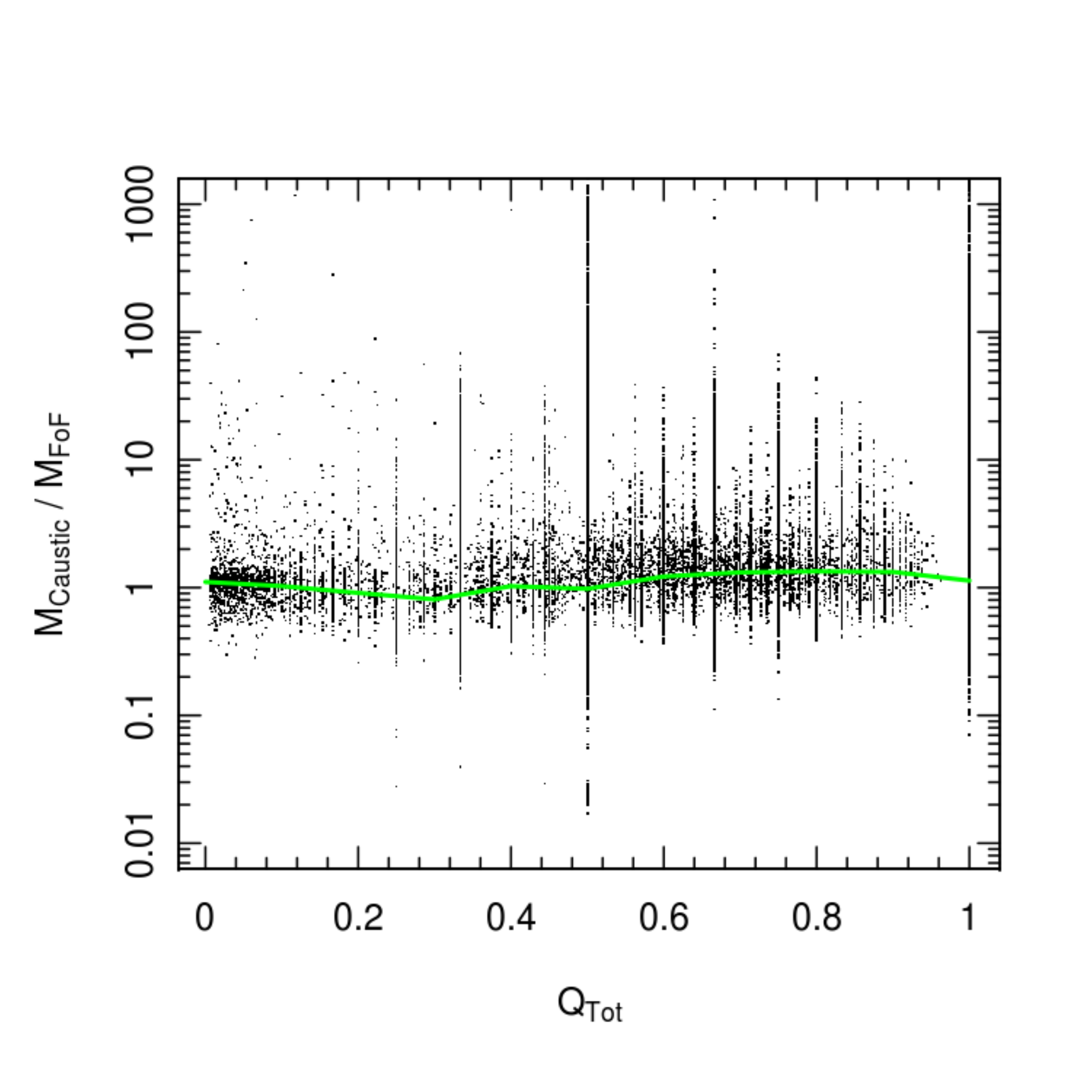}
	\caption{\emph{Left:} A comparison of the performance of the two mass estimation methods as a function of the grouping quality parameter, Q$_{\mathrm{Tot}}$. Black points correspond to caustic masses, and red ones to dynamical masses. Both methods perform worst when Q$_{\mathrm{Tot}}$ is close to 0, but quickly recover at approximately 0.2. The green lines show a rolling median in each 0.1 bin for the caustic mass (solid line) and for the dynamical mass (dashed line). \emph{Right:} The ratio between the caustic and the dynamical mass estimates as a function of Q$_{\mathrm{Tot}}$, with the rolling median in each 0.1 bin overplotted in green.}
	\label{fig:totqplot}
\end{figure*}

One final check is to examine how the caustic mass estimates perform as a function of the quality of the grouping, as defined in \citet{robotham_galaxy_2011}. The total quality parameter, Q$_{\mathrm{Tot}}$, is a measure of how significantly matched individual groups are. The best two-way matching group is defined as being the one which has the largest product for the relative membership between the recovered FoF group and the closest matching mock group (see Eqs. (12) to (14) in \citet{robotham_galaxy_2011}) . In Figure \ref{fig:totqplot} we show how the two mass estimates behave as a function of Q$_{\mathrm{Tot}}$. From the left panel we are able to see that the caustic mass estimates behave in just the same way as the FoF mock dynamical mass estimates, performing well after a total quality factor of about 0.2. This is further demonstrated in the right panel of this figure, where the ratio between the caustic and dynamical mass estimates for the FoF mock groups remains close to unity as a function of Q$_{\mathrm{Tot}}$, with a small tendency for the caustic mass to be systematically greater than the dynamical mass as Q$_{\mathrm{Tot}}$ approaches 1.

Based on these explorations, we chose to create our caustic mass estimates for the mock groups in the catalogue using group centres obtained with the iterative CoL rejection method, and to only include galaxies considered to belong to each group. To generate each set of results, we select the same redshift and multiplicity cuts used in \citet{robotham_galaxy_2011} We calculate the scaling factors $A_{\mathrm{c}}$ for each bin as the necessary value to ensure that the median of the ratio of the mass estimate to the true mass is unity. Values for $A_{\mathrm{c}}$ are listed in Table \ref{table:scalefactor}. 

\subsection{Application to G$^3$Cv1}

With the appropriate scale factors in Table \ref{table:scalefactor}, we run the algorithm on the actual group catalogue itself, using the same redshift and multiplicity bins, the same group centre and only galaxies within each group. This provides us with a full set of halo mass measurements for every group in the catalogue that are complementary of those already existing in the G$^3$Cv1.

\begin{figure*}
	\centering
	\includegraphics[width=1\textwidth]{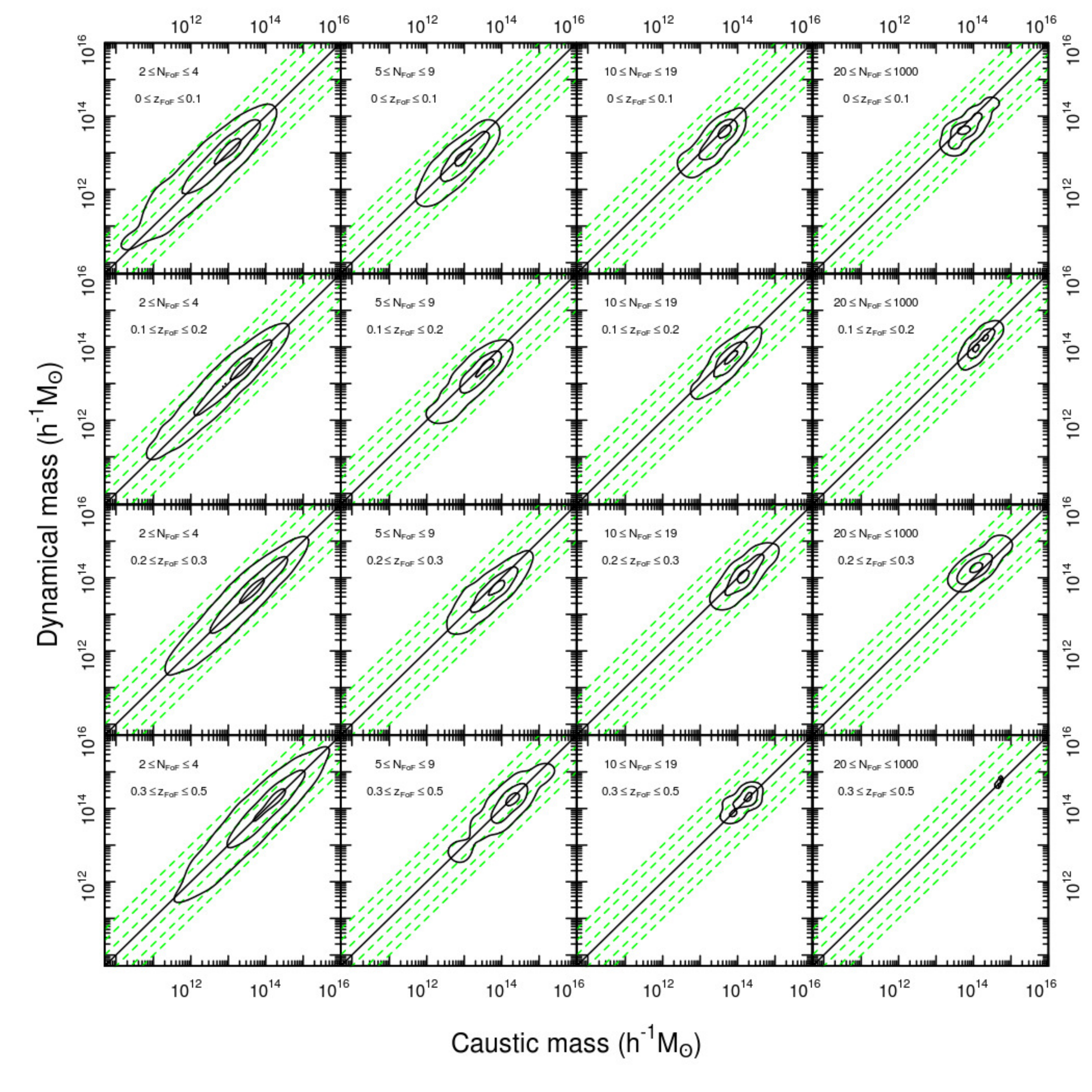}
	\caption{Distribution of the caustic masses compared to dynamical masses for real galaxy groups from G$^3$Cv1. The dashed green lines are as Figure \ref{fig:mockscat_194}, and the solid black lines represent regions containing 10, 50 and 90\% of the data. There is a good correlation between the caustic and dynamical mass estimates as already shown in Fig \ref{fig:mockFoFscat_194} and \ref{fig:mockFoFdens_194} for the mocks.}
	\label{fig:groupscat_194}
\end{figure*}

Both the caustic and the dynamical mass estimates are adjusted by their appropriate scale factors, and then compared. The results of this can be seen in Figure \ref{fig:groupscat_194} for the $r_{\mathrm{AB}} < 19.4$ mag limited sample. By calculating the ratio of the caustic to the dynamical mass, we show that 90.8\% $\pm$ 6.1\% groups have a caustic mass estimate that is within a factor of two of the dynamical mass estimate. In a vast majority of cases the two mass estimates agree very well with each other, particularly for groups at high redshift (bottom row). In all cases the scatter of the data is minimal, with the scatter decreasing with increasing group membership through a combination of better quality mass estimates and the smaller number of galaxies present in these subsets. When interpreting these results, one must bear in mind that the  caustic algorithm has been designed with populous groups in mind; ideally with more than 200 members. The G$^3$Cv1 has exactly one group that fits this criteria, at 264 members. Of 12,200 groups in the G$^3$Cv1, 10,813 of these have between 2 and 4 members and there are only 68 groups that contain more than 20 galaxies. We also observe a disparity between the number of compact groups in the mocks compared to those in GAMA; this is further discussed in \citet{robotham_galaxy_2011}. One also expects small-number statistics to affect the results for these larger groups: this is particularly evident in Table \ref{table:results} which shows the fraction of caustic mass estimates within certain factors of the dynamical mass estimate. As group multiplicity increases these fractions become less reliable, as the number of groups present in each bin drops sharply.

\begin{figure*}
	\centering
	\includegraphics[width=1\textwidth]{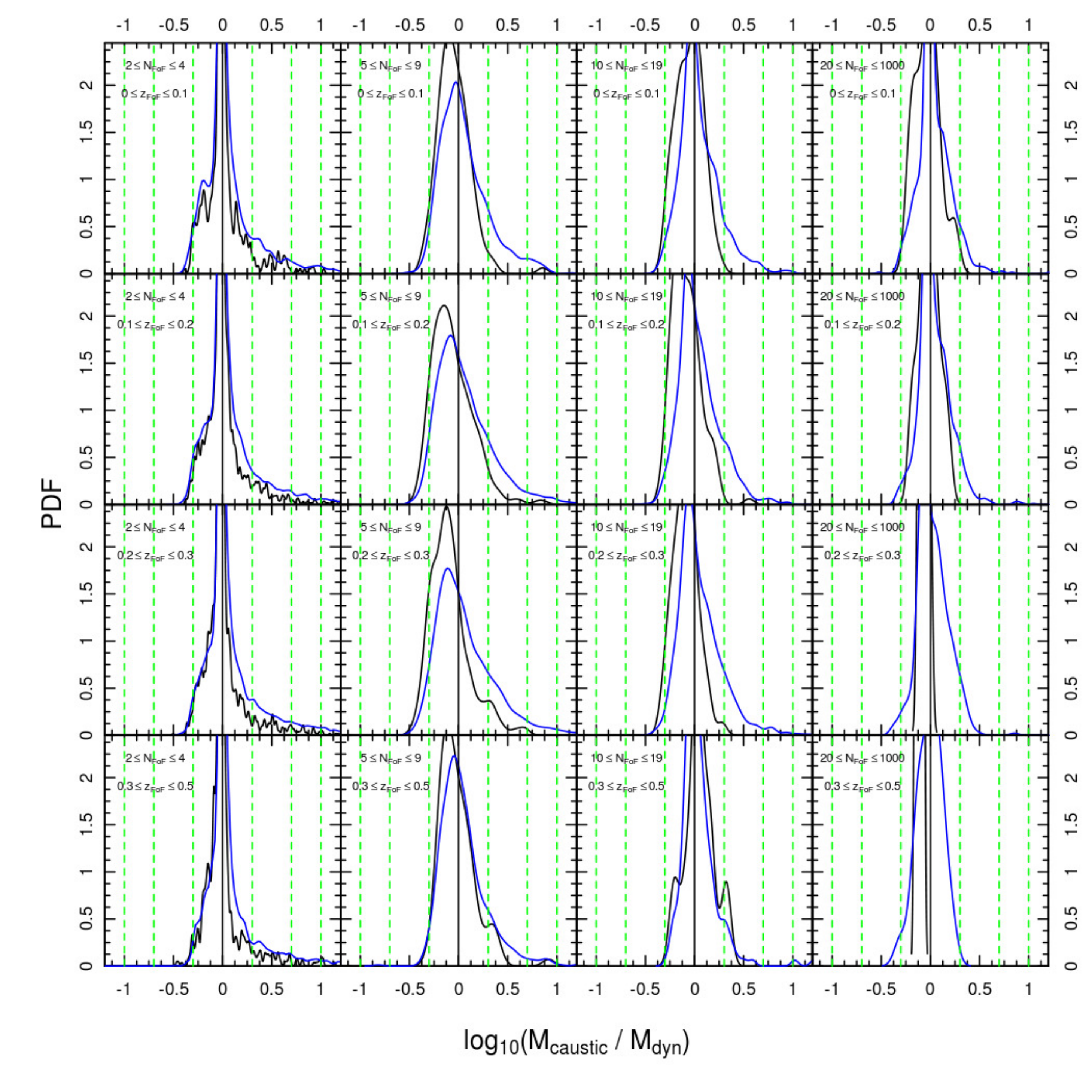}
	\caption{As in Figure \ref{fig:mockdens_194} this figure shows the PDF of the logarithm of the mass ratio between the caustic and dynamical mass estimates for all groups with $\sigma^2 \geq 130$ km$^2$ s$^{-2}$. The blue line shows the same distribution but for the bijectively matched FoF mock catalogues with the same velocity cut to illustrate the smaller scatter between the two methods for the real data.}
	\label{fig:groupdens_194}
\end{figure*}

In order to make a comparison between the groups and mocks as fair as possible, we must account for the fact that velocity uncertainties have not been included in the mocks as this directly affects the kernel, through Eqn. (\ref{eqn:smoothing}). This issue is addressed in \citet{robotham_galaxy_2011}, whereby all groups with $\sigma^2 \geq 130$ km$^2$ s$^{-1}$ are removed from any comparisons between the groups and mocks, arguing that below this cut the velocity dispersion of a group would be significantly affected by this uncertainty and that the dynamical mass estimate is directly proportional to $\sigma^2$. Based on this result, we apply the same velocity cut to the group and mock catalogues. Figures \ref{fig:groupscat_194} and \ref{fig:groupdens_194} display the resulting comparison between the ratio $M_c/M_{\mathrm{dyn}}$ of the caustic and dynamical masses for the groups (black) and mocks (black). There is still some difference between the scatter of the mass estimates for the groups and the mocks and the disagreement is most noticeable in cases where the caustic mass is greater than the dynamical mass. However, if we do not include this velocity cut and include the mock groups with poorly defined masses, the discrepancy between the two distributions increases, with $M_c/M_{\mathrm{dyn}}$ for the groups showing much less scatter than for the mocks. We note that when calculating the mass estimates for the bijectively matched mock groups with $\sigma^2 \geq 130$ km$^2$ s$^{-2}$ for Figure \ref{fig:groupdens_194}, we allow $\sigma_v$ and $\sigma_r$ to \emph{tend} to 0, as Eqn. \ref{eqn:density} diverges if $h_i = 0$ when $h_{\mathrm{opt}} = 0$. In much the same way as done in Table \ref{table:results}, we are able to calculate that the same percentage of caustic and dynamical mass estimates fall within 50\% of the FoF halo mass, implying that both algorithms have correctly been calibrated against the mocks. When comparing the ratios of both mass estimates to the known halo mass for bijectively matched groups, we find that these ratios are virtually the same. From this we can conclude that any final discrepancies in the mass estimates shown in Figures \ref{fig:groupscat_194} and \ref{fig:groupdens_194} must be down to an intrinsic difference between the mocks and the real data, and that both methods are properly calibrated to output results that are median unbiased with respect to the halo masses.

Taking these considerations into account, these new mass estimates appear to agree with the existing dynamical mass estimates for the G$^3$Cv1. While both methods utilise the same velocity and positional information in different forms, the dynamical mass method is not as sensitive to the full 2D velocity profile of each group. This is caused by the dependence of $v$ and $r$ to each other, which is another effect that is considered in the caustic method, whereas the velocity dispersion and group radius used to calculate $M_\mathrm{dyn}$ are treated independently to each other. 

\section{Conclusion}

Using the caustic mass estimation algorithm introduced by DG97 and D99 we have provided complementary caustic mass estimates for the G$^3$Cv1. We have calibrated our implementation of the algorithm by running it on mock GAMA group catalogues where the intrinsic grouping is known; and for bijectively matched friends-of-friends mock groups. This allows us to derive a scaling factor $A_c$ for bins of redshift and multiplicity which provides us with median-unbiased mock mass estimates, and we find that in all cases there is very good agreement between the caustic mass and dynamical mass estimates. Of great note is the tendency for both distributions shown in Figure \ref{fig:mockscat_194} to match each other extremely well, implying that both algorithms perform equally well and equally badly for mock groups: that both estimate methods fail for low multiplicity ($ 2 \leq N \leq 4)$ groups highlights an interesting systematic bias present in both methods that warrants further study. 

Having calibrated the algorithm on the mock catalogues, we apply it to the G3Cv1 and obtain caustic mass estimates for real groups as shown in Figure \ref{fig:groupscat_194}. As with the mocks, we demonstrate that both mass estimates are generally consistent with each other: on average, $90.8 \pm 6.1$ \% of groups have caustic and dynamical mass estimates that agree to within a factor of two. This is strong evidence for the reliability of the caustic mass estimation method, particularly when considering the less than ideal conditions the algorithm faces when working with so many groups of low multiplicity. Despite being designed to work with groups with over 200 members, the caustic method is able to successfully determine accurate mass estimates for groups with as few as two members, highlighting the adaptability and strength of this method when combined with a friends-of-friends algorithm. We stress the importance of applying the caustic mass algorithm alongside a friends-of-friends algorithm; by accurately determining the membership of any given group we reduce the chances that the caustics may be located incorrectly. The tightness of the contours shown in the first column of Figure \ref{fig:groupscat_194} and the large fraction of accurately determined masses given in Table \ref{table:results} are very convincing demonstrations that the caustic method is able to perform well even for very low multiplicity groups.

\begin{table*}
\begin{tabular}{lllllllllllll}
 &\multicolumn{3}{l}{$2 \leq N_{\mathrm{FoF}} \leq 4$}&\multicolumn{3}{l}{$5 \leq N_{\mathrm{FoF}} \leq 9$}&\multicolumn{3}{l}{$10 \leq N_{\mathrm{FoF}} \leq 19$} &\multicolumn{3}{l}{$20 \leq N_{\mathrm{FoF}} \leq 1000$}\\
 &2&5&$\epsilon (N_2)$&2&5&$\epsilon (N_2)$&2&5&$\epsilon (N_2)$&2&5&$\epsilon (N_2)$\\
\hline
$0 \leq z_{\mathrm{FoF}} \leq 0.1$&0.878&0.970&0.010&0.862&0.983&0.030&0.958&1.0&0.030&0.769&1.0&0.152\\
$0.1 \leq z_{\mathrm{FoF}} \leq 0.2$&0.898&0.977&0.005&0.836&0.904&0.020&0.949&1.0&0.020&0.978&1.0&0.022\\
$0.2 \leq z_{\mathrm{FoF}} \leq 0.3$&0.904&0.979&0.006&0.794&0.997&0.028&0.952&1.0&0.028&1.0&1.0&0\\
$0.3\leq z_{\mathrm{FoF}} \leq 0.5$&0.909&0.978&0.009&0.904&0.979&0.034&0.889&1.0&0.118&1.0&1.0&0\\
\end{tabular}
\caption{Fraction of groups that have a caustic mass estimate that is within a factor of 2 and 5 from the dynamical mass estimate. The error $\epsilon (N_2)$ is defined as $\sqrt{1/N_2 - 1/N_{\mathrm{tot}}}$ where $N_2$ is the number of groups within a factor 2 and $N_{\mathrm{tot}}$ is the total number of groups for a given bin. The number of groups present in each multiplicity bin drops sharply (from thousands of groups to 10 or less) after the first multiplicity bin, drastically lowering the relevance of these statistics.}
\label{table:results}
\end{table*}

We wish to draw particular attention to the result shown in Figure \ref{fig:groupdens_194} that the ratio between the caustic and dynamical mass estimates in the groups is very sensitive to the application of a velocity cut at $\sigma^2 \geq 130$ km$^2$ s$^{-2}$, as required by the fact that the mocks (and gene the mock group catalogues) have not been analysed with realistic velocity errors. By discarding groups whose velocity dispersions would be badly affected by the inclusion of velocity errors, we ensure that we apply a fair comparison to the group catalogue, which does include velocity errors (and these velocity errors play a crucial part in the calculation of the kernel on which the caustic algorithm depends). Once this cut is applied, we show that the distribution of ratio between the caustic and dynamical masses is consistent between the mocks and groups, particularly for groups with 2 to 4 members. 

These caustic mass estimates serve to add further credibility to the dynamical mass estimates, as the caustic algorithm is sensitive to the escape velocity profile of a given group out to its full extent. That both dynamical mass and caustic mass estimates for the mocks and the groups give such similar results despite their relative independence from each other is a positive result that not only reinforces the dynamical mass estimates, but also demonstrates the power of the caustic approach when combined with a friends-of-friends algorithm. 

The G$^3$Cv1 will be made publicly available on the GAMA website as and when the associated redshift data is made available. Those wishing to use the group catalogue before this time should contact ASGR at asgr@st-and.ac.uk. The caustic mass estimation algorithm developed by the author has been written in R \citep{Rproject}. A non-GAMA specific version of the algorithm is freely available to use and can be downloaded from http://www.gama-survey.org/pubs/.

\section{Acknowledgements}

The authors wish to thank Antonaldo Diaferio for his helpful private communications during the writing of this paper. MA would like to acknowledge funding from the University of St Andrews and the International Centre for Radio Astronomy. PN acknowledges a Royal Society URF and an ERC StG grant (DEGAS-259586). AM is supported by funding from the Science \& Technologies Funding Council.

GAMA is a joint European-Australasian project based around a spectroscopic campaign using the Anglo-Australian Telescope. The GAMA input catalogue is based on data taken from the Sloan Digital Sky Survey and the UKIRT Infrared Deep Sky Survey. Complementary imaging of the GAMA regions is being obtained by a number of independent survey programs including GALEX MIS, VST KIDS, VISTA VIKING, WISE, Herschel-ATLAS, GMRT and ASKAP providing UV to radio coverage. GAMA is funded by the STFC (UK), the ARC (Australia), the AAO, and the participating institutions. The GAMA website is http://www.gama-survey.org/.

\footnotesize
\bibliographystyle{mn2e}
\setlength{\bibhang}{2.0em}
\setlength{\labelwidth}{0.0em}

\normalsize

\label{lastpage}

\end{document}